\documentclass[sigconf,authorversion,nonacm]{acmart}
\makeatletter
\def\@ACM@checkaffil{
    \if@ACM@instpresent\else
    \ClassWarningNoLine{\@classname}{No institution present for an affiliation}%
    \fi
    \if@ACM@citypresent\else
    \ClassWarningNoLine{\@classname}{No city present for an affiliation}%
    \fi
    \if@ACM@countrypresent\else
        \ClassWarningNoLine{\@classname}{No country present for an affiliation}%
    \fi
}

\usepackage{geometry}

\usepackage{balance}
\usepackage{bm}
\usepackage{threeparttable}
\usepackage{booktabs}
\usepackage{multirow}
\usepackage{enumitem}
\usepackage{balance}
\usepackage{diagbox}
\usepackage{enumitem}
\setlist[itemize]{leftmargin=*}

\usepackage{amsmath} 
 
\usepackage{amssymb}
\usepackage{pdfrender}
\usepackage{graphicx}
\usepackage{caption}
\usepackage{subcaption}

\usepackage[hang,flushmargin]{footmisc}

\makeatother

\newcommand{\dope}{\centering\arraybackslash}

\begin{document}
\title[Scaling New Frontiers: Insights into Large Recommendation Models]{Scaling New Frontiers: \\Insights into Large Recommendation Models}

\author{Wei Guo$^{1\dagger}$, Hao Wang$^{2\dagger}$, Luankang Zhang$^{2\dagger}$, Jin Yao Chin$^{1\dagger}$, Zhongzhou Liu$^{1}$, Kai Cheng$^{2}$, \\
Qiushi Pan$^{2}$, Yi Quan Lee$^{1}$, Wanqi Xue$^{1}$, Tingjia Shen$^{2}$, Kenan Song$^{1}$, Kefan Wang$^{2}$,
Wenjia Xie$^{2}$, \\ Yuyang Ye$^{2}$,
Huifeng Guo$^{1}$, Yong Liu$^{1}$*, Defu Lian$^{2}$*, Ruiming Tang$^{1}$*, Enhong Chen$^{2}$*}

\affiliation{
  \institution{$^{1}$Huawei Noah's Ark Lab}
  \{guowei67, chin.jin.yao, liu.yong6, tangruiming\}@huawei.com
}

\affiliation{
  \institution{$^{2}$University of Science and Technology of China}
  \{wanghao3, liandefu, cheneh\}@ustc.edu.cn \\
   \{zhanglk5\}@mail.ustc.edu.cn
}

\renewcommand{\shortauthors}{Guo and Wang, et al.}

\begin{abstract}

Recommendation systems are essential for filtering data and retrieving relevant information across various applications. Recent advancements have seen these systems incorporate increasingly large embedding tables, scaling up to tens of terabytes for industrial use. However, the expansion of network parameters in traditional recommendation models has plateaued at tens of millions, limiting further benefits from increased embedding parameters. Inspired by the success of large language models (LLMs), a new approach has emerged that scales network parameters using innovative structures, enabling continued performance improvements.
A significant development in this area is Meta's generative recommendation model HSTU, which illustrates the scaling laws of recommendation systems by expanding parameters to thousands of billions. This new paradigm has achieved substantial performance gains in online experiments. 
In this paper, we aim to enhance the understanding of scaling laws by conducting comprehensive evaluations of large recommendation models.
Firstly, we investigate the scaling laws across different backbone architectures of the large recommendation models. Secondly, we conduct comprehensive ablation studies to explore the origins of these scaling laws. We then further assess the performance of HSTU, as the representative of large recommendation models, on complex user behavior modeling tasks to evaluate its applicability. Notably, we also analyze its effectiveness in ranking tasks for the first time. Finally, we offer insights into future directions for large recommendation models.
Supplementary materials for our research are available on GitHub at \textcolor{blue}{\url{https://github.com/USTC-StarTeam/Large-Recommendation-Models}}.

\end{abstract}

\keywords{Scaling Law, Large Recommendation Model, Generative Recommendation}
\maketitle

\let\thefootnote\relax\footnote{{$\dagger$ Equal contribution.}}
\let\thefootnote\relax\footnote{{* 
 Corresponding authors.}}

\vspace{-20pt}
\section{Introduction}
In the current era characterized by an overwhelming influx of information, web services such as TikTok, Taobao, and YouTube are inundated with vast amounts of data. To effectively navigate this deluge and recommend relevant items to users, real-world recommendation systems must efficiently identify items from a pool of millions to billions of candidates. This is typically achieved through a multi-stage framework, which involves sequential processes of recall, ranking, and re-ranking.

In the ongoing quest for improved user experiences and increased platform revenues, the scalability of models within industrial recommendation systems has become a focal point of research and development. Early approaches sought to enhance scalability by expanding the sparse parameters, such as embedding tables, through the integration of additional categorical and cross features. This expansion can lead to models with billions or even trillions~\cite{zhao2020distributed, chang2023pepnet} of features, resulting in embedding parameters that require hundreds of gigabytes to terabytes of storage in large-scale applications. An alternative method involves increasing the embedding dimension, as demonstrated by multi-embedding models~\cite{guo2023embedding}, which aim to address embedding collapse and enhance scalability. However, recent research~\cite{zhai2024actions} indicates that simply enlarging embedding tables does not effectively improve model capacity and is computationally inefficient. Consequently, it is imperative to explore new perspectives for addressing scalability challenges.

Drawing inspiration from the remarkable success of large language models (LLMs), recent research has increasingly focused on scaling up dense parameters in recommendation systems by developing innovative structures that enable sustained performance growth through layer stacking. From the perspective of feature interaction modeling, Wukong~\cite{zhang2024wukong} investigates the scaling law by refining the feature interaction module. This is achieved through an effective network architecture that combines stacked factorization machines with linear blocks, aiming to facilitate loss scaling as dense parameters increase. However, while the authors assert the presence of a scaling law, the results show only modest reductions in loss curves, and the improvements on some public datasets are relatively minor, limiting the work's impact and attention.
Conversely, in another perspective of generative recommendation (GR) using user behavior sequences as input, HSTU~\cite{zhai2024actions} introduces an innovative transformer-based structure. This model replaces the Softmax function with the SiLU activation function and incorporates additional multiplicative terms within the self-attention module. It demonstrates empirical scalability in line with a power-law relationship concerning training compute, spanning three orders of magnitude and achieving a scale comparable to large language models like GPT-3 and LLaMa-2.

While recent research has extensively explored the use of large language models (LLMs), such as ChatGPT, to enhance recommendations (referred to as \textbf{\textsl{LLMs enhanced recommendation}}) through their world knowledge and advanced logical reasoning capabilities, the challenge of scaling up the dense parameters of recommendation models (referred to as \textbf{\textsl{large recommendation models}}) remains underexplored.

\begin{definition}[Large Recommendation Model]
A large recommendation model is a scalable system designed to process and analyze multi-modal and heterogeneous data. It supports a wide range of recommendation tasks and enhances performance by leveraging increased model parameters and larger datasets.
\end{definition}

As we stand at the intersection of current and next-generation recommendation technologies, our focus shifts towards the relatively unexplored domain of large recommendation models across various tasks and the underlying principles of their scaling laws.
Given the significant business improvements and powerful scaling effects demonstrated by the GR paradigm, this paper presents a comprehensive analysis of the scalability factors in current large recommendation models. Specifically, we aim to deepen the understanding of the scaling law of large recommendation models by evaluating their capacities in complex behavior modeling and ranking tasks, thereby uncovering their potential in a wider range of downstream applications.

In summary, our contributions are fourfold, specifically emphasizing the most critical and emerging characteristics of scalable large recommendation models:

\begin{itemize}[leftmargin=*,align=left]
\item We analyze the scalability of various popular transformer-based architectures for large recommendation models, including HSTU, Llama, GPT, and SASRec, by evaluating their performance with an increasing number of attention blocks.
\item We conduct comprehensive ablation studies and parameter analysis on HSTU, as the representative of large recommendation models, to explore the origins of its scaling law. Additionally, we enhance the scalability of SASRec, a legacy transformer-based sequential recommendation model, by integrating effective modules from scalable large recommendation models.
\item We further investigate the performance of HSTU on complex user behavioral sequence data, identifying areas for improvement in modeling intricate user behaviors, particularly with data involving side information and multiple behaviors.
\item To the best of our knowledge, we are the first to thoroughly evaluate HSTU on ranking tasks, demonstrating their scalability in this context. Our evaluations also provide insights into designing effective large recommendation models for ranking, considering datasets and hyper-parameters.
\end{itemize}

\begin{figure}
    \centering
    \includegraphics[width=1.0\linewidth]{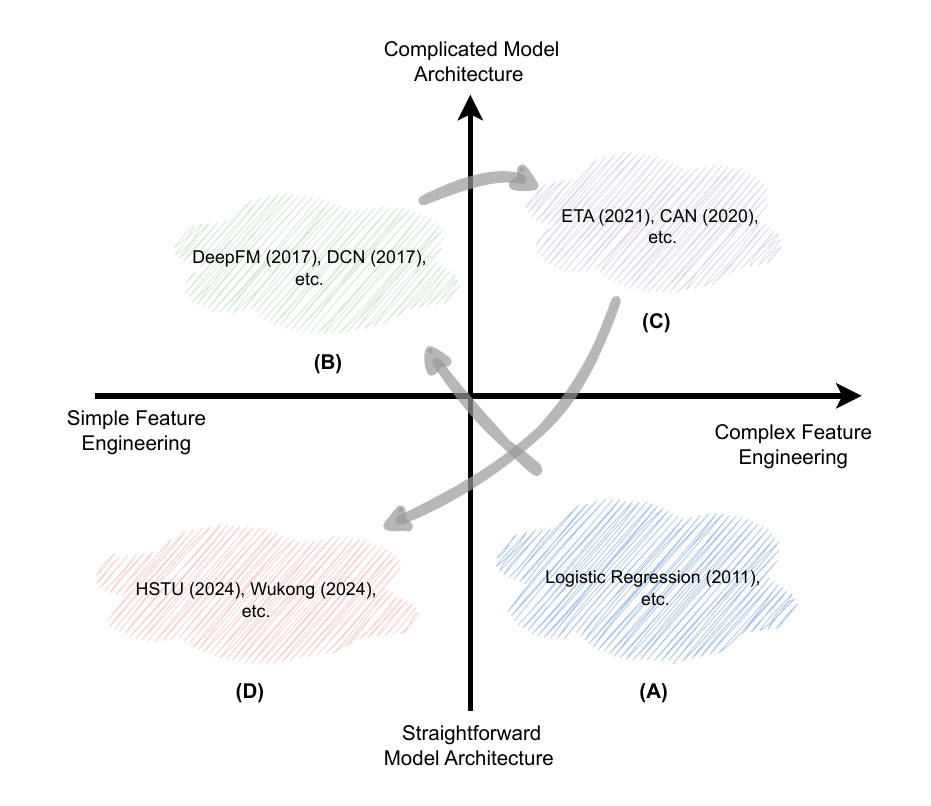}
    \caption{Illustration of the shift in modeling paradigms.}
    \label{fig:shift_paradigm}
        \vspace{-4mm}
\end{figure}

\begin{figure*}
    \centering
    \includegraphics[width=0.98\textwidth]{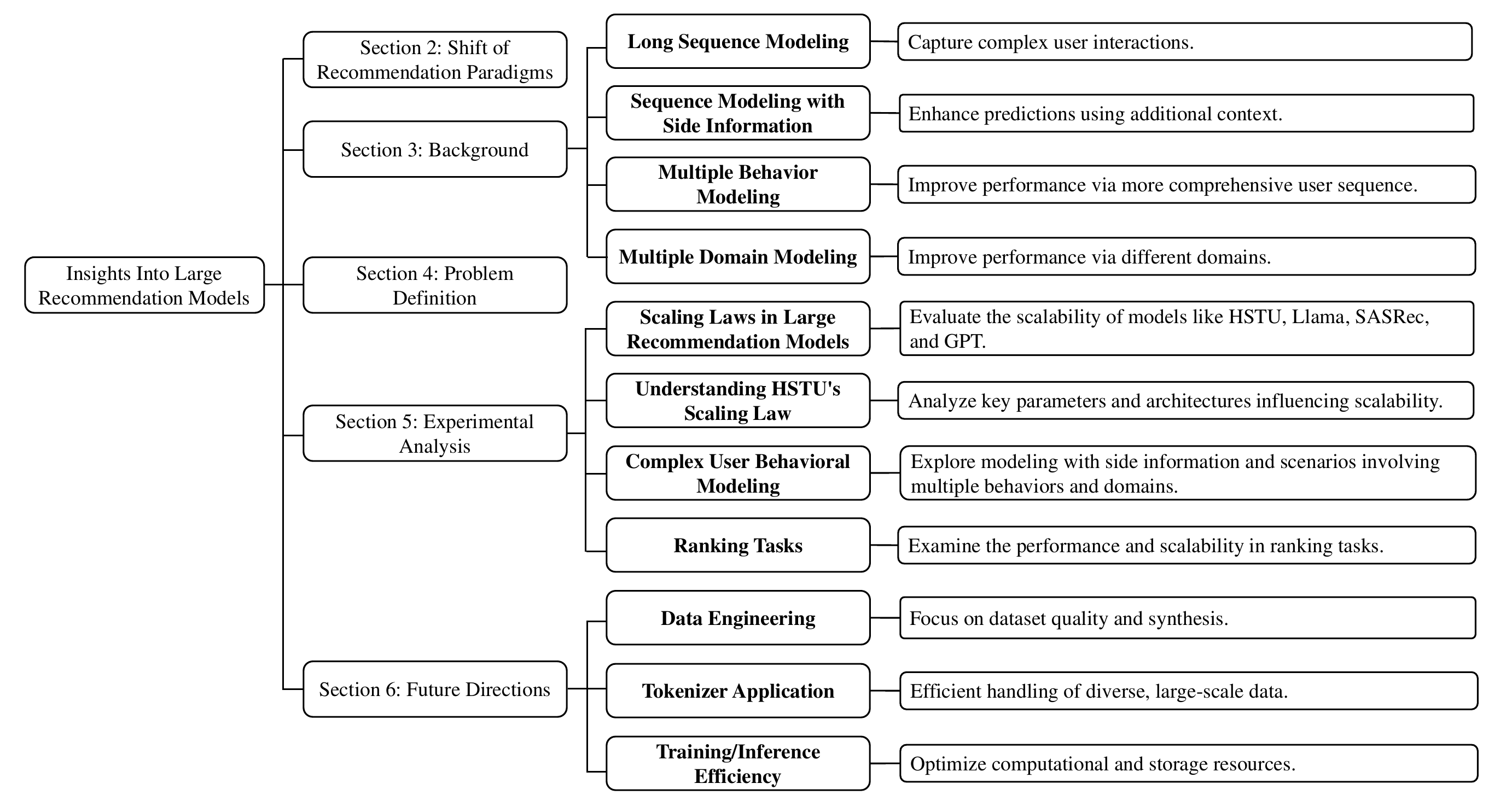}
    \caption{The figure outlines the structure of the paper, starting with the shift in user behavior modeling paradigms discussed in Section~\ref{sec:observation}, followed by advancements in the field in Section~\ref{sec:background}. It then covers the problem definition in Section~\ref{sec:problem-definition}, presents preliminary experimental results in Section~\ref{sec:experiments}, and concludes with future research directions in Section~\ref{sec:future-directions}.}
    \label{fig:intro_section_pdf}
    
\end{figure*}

This paper is organized as follows: Section~\ref{sec:observation} provides a comprehensive review of the shift in user behavior modeling paradigms, particularly within large recommendation models, highlighting key changes and trends, as illustrated in Figure~\ref{fig:shift_paradigm}. Section~\ref{sec:background} presents the latest advancements in the related field. In Section~\ref{sec:problem-definition}, we define the problem under investigation. Preliminary experimental results related to the research problems are discussed in Section~\ref{sec:experiments}. Section~\ref{sec:future-directions} explores prospects for future research by proposing potential directions. Finally, Section~\ref{sec:conclusion} summarizes our key findings and contributions. The overall framework is shown in Figure~\ref{fig:intro_section_pdf}.

\section{Shift of Modeling Paradigm}
\label{sec:observation}

With the emergence of \textbf{\textsl{large recommendation models}}, the focus of recommendation systems  is undergoing significant transformations. As shown in the Figure~\ref{fig:shift_paradigm},
a pivotal shift is the reduced emphasis on feature engineering and model architecture design. 

Initially, the development of recommendation systems is heavily influenced by constraints in computational resources, prompting researchers to concentrate on crafting effective features and utilizing simple predictive models~\cite{zhang2021deep, guo2021dual}, as depicted in Figure~\ref{fig:shift_paradigm}(A). 
A prime example of this approach is Logistic Regression (LR) \cite{mcmahan2011follow}, which assigns adaptive weights to various features to improve prediction accuracy. These features typically include user attributes (such as country, gender, and age), item characteristics (such as brand and category), and contextual elements (including weather, hour of the day, and day of the week).

Later, with the rise of deep learning, there has been a significant shift towards developing more complex models that fully leverage the parallel computing capabilities of GPUs, as depicted in Figure~\ref{fig:shift_paradigm}(B).
One notable example is DeepFM \cite{guo2017deepfm}, which introduces a hybrid architecture combining a shallow FM model and a deep DNN model to simultaneously learn low-order and high-order feature interactions.
Similarly, the Deep \& Cross Network (DCN) \cite{wang2017deep} explicitly applies feature crossing at each layer, allowing feature interaction orders to increase in a layer-wise manner.

While the performance improvements from simply designing more complex deep models have reached a plateau, a recent trend in recommendation systems is to revisit feature engineering through learnable methods, as illustrated in Figure~\ref{fig:shift_paradigm}(C). For instance, the ETA model \cite{ETA} leverages locality-sensitive hashing and Hamming distance to compute item similarity, allowing for the selection of the most significant features from long sequences to improve recommendation accuracy. Similarly, CAN \cite{zhou2020can} employs a meta-network to approximate the Cartesian product, effectively modeling the cross-relations between two features.

In the current era, the extraordinary success of large recommendation models, combined with the widespread acceptance of scaling laws, indicates that computational power will be crucial for future improvements in model performance, as depicted in Figure~\ref{fig:shift_paradigm}(D). 
The scaling law, which demonstrates a power-law relationship between model loss and key variables such as model size, dataset size, and computational resources, shapes our vision for the evolution of large-scale recommendation model development.
Looking ahead, we anticipate two important directions: firstly, expanding the dataset by more effectively mining and leveraging user behavior sequences across various domains for user life-cycle modeling; and secondly, scaling up the model size while ensuring training and inference under certain cost constraints with efficiency optimization.
\section{Background}
\label{sec:background}

Large recommendation models based on the GR paradigm can be viewed as sophisticated user behavior models. These models process user behavior sequences as input and learn behavior dependencies through transductive learning. In this section, we review key research directions in user behavior modeling to provide a comprehensive understanding of this field \cite{he2023survey, liu2023user,wu2024survey}.

\paragraph{Long Sequence Modeling}
The rapid development of online services has led to the accumulation of extensive user behavioral data. Consequently, modeling long sequences of user behavior is crucial for industrial recommendation systems. This focus has driven significant efforts in the long sequence modeling to effectively capture the multifaceted and evolving interests of users. Research in this area has evolved through three phases: the initial use of memory networks \cite{chen2018sequential, MIMN, ren2019lifelong}, followed by advancements in user behavior recall techniques \cite{qin2023learning, SIM, xie2024bridging, xie2024breaking,xuxiang}, and most recently, the emergence of efficient transformer models \cite{pancha2022pinnerformer, wu2021linear, yu2024ifa}.

\paragraph{Sequence Modeling with Side Information}
Beyond the primary user behavior sequence, which includes the user's historical interactions, side information such as temporal data and user/item attributes can enhance sequence modeling. For temporal information, the simplest and most frequently used approach is to arrange the user's historical interactions in chronological order \cite{kang2018selfattentive, fei2019bert4rec,yin2023apgl4sr,wang2021hypersorec,han2023guesr}. Beyond ordering, the time intervals between interactions provide insights into user preferences \cite{li2020time, ye2020time, cho2020meantime, zhai2024actions,wang2021decoupled,wang2019mcne}. Regarding user/item attributes, these can offer additional context to the user behavior modeling process. User features, such as age, gender, and occupation, and item features, such as price, category, and brand, can serve as complementary auxiliary information. However, privacy concerns and regulations like GDPR may limit access to user features, leading models to primarily focus on item attributes.

\paragraph{Multiple Behavior Modeling}
Traditional recommendation systems often focus on a single type of user behavior. However, real-world user interactions are multifaceted, including actions such as clicks, shares, and purchases. Leveraging this multi-behavioral data is essential for constructing a comprehensive user representation, thereby enabling more precise recommendations. This challenge has led to the emergence of the Multi-Behavior Sequential Recommendation (MBSR) problem \cite{meng2023coarse,meng2023hierarchical,guo2023compressed,wang2024denoising,han2024efficient}. Current MBSR methodologies can be broadly categorized into two main approaches: (1) segmenting item sequences into subsequences according to behavior categories, modeling each subsequence independently, and subsequently integrating them for prediction~\cite{guo2019buying, gu2020deep, xia2020multiplex, wu2022multiview}; and (2) modeling the entire item sequence while incorporating behavior types as auxiliary inputs~\cite{gao2019neural, yuan2022multibehavior, su2023personalized, rajput2024recommender, zhai2024actions, liu2024multi}.

\paragraph{Multiple Domain Modeling}
In addition to integrating interactions from various user behaviors, several methods leverage auxiliary domain interactions to enrich user behavior profiles, thereby improving recommendation precision in the target domain. This strategy is commonly known as cross-domain or multi-domain sequential recommendation~\cite{Cao2022contrastive, ma2024triple, li2022recguru, tong2024mdap, shen2024exploring,yin2024learning,zhang2024unified}. Recent advancements~\cite{zhou2020s3rec, hou2022towards, li2023text, hou2023learning, yin2024learning,wangmf} have incorporated auxiliary information, such as product descriptions, titles, and brands, which act as semantic bridges across domains. These approaches typically follow a two-stage framework: initially, they employ pre-training tasks to develop enhanced universal representations using the auxiliary information; subsequently, they fine-tune the model within a single domain to enable effective adaptation to new scenarios.
\section{Problem Definition}
\label{sec:problem-definition}

In this study, we aim to present a comprehensive analysis of the scalability sources of current large recommendation models and reveal the capacity of current large recommendation models in complex behavior modeling and ranking tasks.
We define two primary sets: the user set \(\mathcal{U} = \{u_1, u_2, \ldots, u_{|\mathcal{U}|}\}\) and the item set \(\mathcal{V} = \{v_1, v_2, \ldots, v_{|\mathcal{V}|}\}\), where \(|\mathcal{U}|\) and \(|\mathcal{V}|\) represent the total number of users and items, respectively. 

For each user \(u \in \mathcal{U}\), we model their interactions as a sequence \(X_u = \{x_1, x_2, \ldots, x_n\}\), where each \(x_i \in \mathcal{V}\) is an item the user has interacted with, listed in chronological order. The sequence length is capped at \(n\); shorter sequences are padded, and longer sequences are truncated to retain the most recent interactions. Our goal is to train a model that predicts the next item \(x_{n+1}\) a user will interact with, based solely on the interaction sequence \(X_u\).

For simplicity, we divide the multi-stage recommendation process into recall and ranking, omitting other possible steps like pre-ranking and re-ranking, which we will explore in future work.

\textbf{Recall:} In single-behavior recall, given a user's historical interaction sequence \(X_u\), the recall model is trained to select items from the entire item set \(\mathcal{V}\) that the user might interact with at step \(n+1\), forming a candidate set \(I\). 

\textbf{Ranking:} For the items in the candidate set \(I\) identified by the recall model, the ranking model evaluates them based on the user's historical interaction sequence \(X_u\). 
It predicts the user's preference for each item, such as the probability of a click, and sorts the items to determine the most relevant ones to display to the user.

To further improve recommendation performance, we incorporate side information into the generative framework.

\textbf{Recommendation with Side Information:} Let \( C \) denote the set of side information attributes, such as time, location, and user age. For each user \( u \in \mathcal{U} \), the interaction sequence is represented as \( X_u = \{(x_1, c_1), (x_2, c_2), \ldots, (x_n, c_n)\} \), where \( x_i \in \mathcal{V} \) is the item interacted with at the \( i \)-th step, and \( c_i \in C \) is the relevant side information available at that step. The goal is to train a model that leverages both the interaction sequence \( X_u \) and the associated side information to predict the next item \( x_{n+1} \) that the user will interact with in the subsequent time step.

Additionally, we address complex scenarios by exploring multi-behavior and multi-domain recommendations to capture intricate user interactions and cross-domain influences.

\textbf{Multi-behavior Recommendation:} Let \(\mathcal{B} = \{b_1, b_2, \ldots, b_{|\mathcal{B}|}\}\) denote the set of different user behaviors. In multi-behavior recommendation, the interaction sequence is constructed by pairing each user interaction with its corresponding behavior, represented as \(X_u = \{(x_1, b_1), (x_2, b_2), \ldots, (x_n, b_n)\}\). Here, \(x_i\) denotes the item, and \(b_i\) represents the behavior associated with that item.
The model's objective is to predict the next item \(x_{n+1}\) based on \(X_u\).

\textbf{Multi-domain Recommendation:} In multi-domain recommendation, we consider multiple domains denoted as \(D^i\), where \(i = 1, 2, \ldots, d\) and \(d\) represents the total number of domains. Each domain \(D^i\) comprises a user set \(\mathcal{U}^i\), an item set \(\mathcal{V}^i\), and an interaction set \(X_{\mathcal{U}}^i\). The objective is to utilize the combined cross-domain interaction sequences \(X = X_{\mathcal{U}}^1 \cup X_{\mathcal{U}}^2 \cup \ldots \cup X_{\mathcal{U}}^d\) to predict the next item \(x_{n+1}^i \in \mathcal{V}^i\) that a user will interact with in a specific domain \(i\).

By defining these problems, we establish a comprehensive framework to address the entire process of recall and ranking in recommendation systems. This framework incorporates side information and extends the scope from single-behavior to complex multi-domain and multi-behavior scenarios, thereby enhancing the adaptability and accuracy of recommendation models.

\section{EXPERIMENTS}
\label{sec:experiments}

In this section, we conduct extensive experiments on multiple datasets to address the following research questions.
\begin{itemize}[leftmargin=*,align=left]
\item \textbf{(RQ1)} How does model depth influence the scaling laws of large recommendation models? 

\item \textbf{(RQ2)} Where does the scaling law of large recommendation models originate from?

\item \textbf{(RQ3)} What is the performance of HSTU when applied to complex user behavioral sequence modeling?

\item \textbf{(RQ4)} What is the performance of HSTU on ranking task? What are the key points to obtain the scaling law?
\end{itemize}

\begin{figure}
    \centering
    \includegraphics[width=0.9\linewidth]{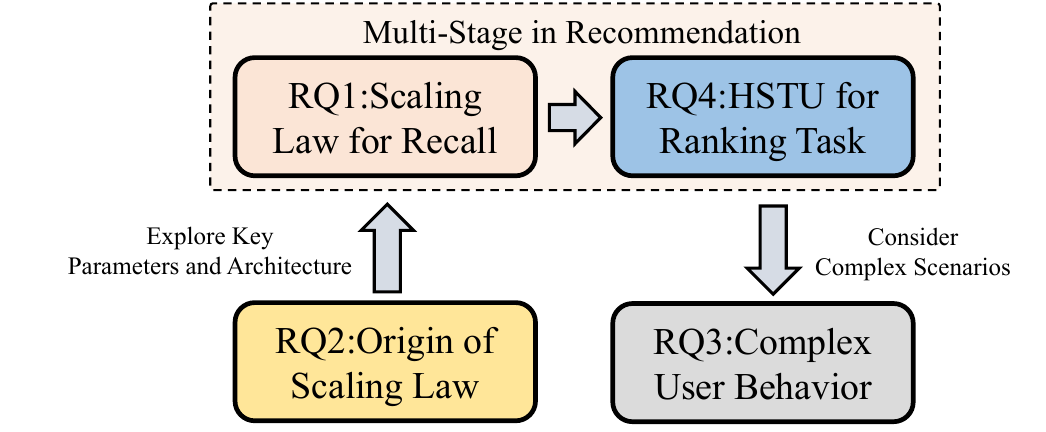}
    \caption{The research questions addressed in experiments.}
    \label{fig:intro_fig_pdf}
    \vspace{-4mm}
\end{figure}

As depicted in Figure~\ref{fig:intro_fig_pdf}, we first introduce the experimental settings in Section~\ref{sec:experimental-settings}. Then, we present an overview of model performances and the scalability of some popular backbones in Section~\ref{sec:comparison-all} to answer RQ1. 
As the representative of large recommendation models, we next conduct an in-depth analysis of HSTU and address RQ2 in Section~\ref{sec:ablation}. Besides, we also conduct comprehensive evaluations of HSTU on complex user behavioral sequence modeling and ranking tasks in Sections~\ref{sec:complex-sequence-modeling} and~\ref{sec:ranking-task}, respectively, to address RQ3 and RQ4.

\subsection{Experimental Settings}
\label{sec:experimental-settings}

\begin{table*}[h!]
    \centering
\begin{tabular}{@{}c c >{\dope}p{1.3cm} >{\dope}p{1.3cm} >{\dope}p{1.3cm}>{\dope}p{1.3cm} >{\dope}p{1.3cm} >{\dope}p{1.3cm} >{\dope}p{1.3cm} >{\dope}p{1.3cm} >{\dope}p{1.3cm}@{}}
    \toprule
    \multirow{2}{*}{\textbf{Model}} & \multirow{2}{*}{\textbf{\# Blocks}} & \multicolumn{3}{c}{\textbf{ML-1M}} & \multicolumn{3}{c}{\textbf{ML-20M}} & \multicolumn{3}{c}{\textbf{AMZ-Books}} \\
    & & \textbf{HR@10} & \textbf{NDCG@10} & \textbf{MRR} & \textbf{HR@10} & \textbf{NDCG@10} & \textbf{MRR} & \textbf{HR@10} & \textbf{NDCG@10} & \textbf{MRR} \\
    \midrule
    \multirow{5}{*}{HSTU} 
    & 2  & 0.2923 & 0.1628 & 0.1395 & 0.2915 & 0.1642 & 0.1409 & 0.0564 & 0.0308 & 0.0279 \\
    & 4  & 0.3115 & 0.1778 & 0.1529 & 0.3218 & 0.1849 & 0.1582 & 0.0617 & 0.0338 & 0.0305 \\
    & 8  & 0.3299 & 0.1857 & 0.1572 & 0.3403 & 0.1990 & 0.1709 & 0.0649 & 0.0357 & 0.0322 \\
    & 16 & 0.3322 & 0.1887 & 0.1601 & 0.3520 & 0.2079 & 0.1787 & 0.0680 & 0.0377 & 0.0340 \\
    & 32 & 0.3298 & 0.1863 & 0.1580 & 0.3569 & 0.2113 & 0.1814 & 0.0584 & 0.0325 & 0.0295 \\
    \midrule
    \multirow{5}{*}{Llama} 
    & 2  & 0.3029 & 0.1697 & 0.1450 & 0.3044 & 0.1724 & 0.1475 & 0.0510 & 0.0275 & 0.0252 \\
    & 4  & 0.3153 & 0.1796 & 0.1539 & 0.3277 & 0.1887 & 0.1615 & 0.0537 & 0.0292 & 0.0266 \\
    & 8  & 0.3232 & 0.1848 & 0.1583 & 0.3449 & 0.2008 & 0.1718 & 0.0547 & 0.0296 & 0.0269 \\
    & 16 & 0.3298 & 0.1872 & 0.1594 & 0.3495 & 0.2055 & 0.1764 & 0.0227 & 0.0117 & 0.0112 \\
    & 32 & 0.3299 & 0.1896 & 0.1626 & 0.3551 & 0.2090 & 0.1791 & 0.0210 & 0.0110 & 0.0107 \\
    \midrule
    \multirow{5}{*}{GPT} 
    & 2  & 0.2798 & 0.1564 & 0.1343 & 0.2419 & 0.1333 & 0.1155 & 0.0568 & 0.0307 & 0.0279 \\
    & 4  & 0.2803 & 0.1543 & 0.1319 & 0.0284 & 0.0148 & 0.0162 & 0.0356 & 0.0191 & 0.0180 \\
    & 8  & 0.0353 & 0.0162 & 0.0178 & 0.0302 & 0.0147 & 0.0151 & 0.0049 & 0.0026 & 0.0032 \\
    & 16 & 0.0270 & 0.0133 & 0.0162 & 0.0264 & 0.0127 & 0.0138 & 0.0050 & 0.0026 & 0.0032 \\
    & 32 & 0.0247 & 0.0115 & 0.0140 & 0.0312 & 0.0145 & 0.0150 & 0.0058 & 0.0029 & 0.0033 \\
    \midrule
    \multirow{5}{*}{SASRec} 
    & 2  & 0.2824 & 0.1594 & 0.1375 & 0.2781 & 0.1553 & 0.1330 & 0.0561 & 0.0305 & 0.0276  \\
    & 4  & 0.2744 & 0.1543 & 0.1335 & 0.0599 & 0.0294 & 0.0284 & 0.0544 & 0.0300 & 0.0272 \\
    & 8  & 0.2183 & 0.1186 & 0.1030  & 0.0326 & 0.0156 & 0.0169 & 0.0084 & 0.0042 & 0.0043 \\
    & 16 & 0.0431 & 0.0184 & 0.0176 & 0.0349 & 0.0167 & 0.0177  & 0.0095 & 0.0044 & 0.0042\\
    & 32 & 0.0366 & 0.0181 & 0.0195 & 0.0301  & 0.0159 & 0.0169  & 0.0084 & 0.0044 & 0.0045\\
    \bottomrule
    \end{tabular}
    \caption{Performance comparison of different backbones.}
    \vspace{-5mm}
    \label{tab:backbone_retrieval}
\end{table*}

\subsubsection{Datasets}
To assess the effectiveness of the recommendation models, we conduct experiments using both single-behavior and multi-behavior datasets.
For the single-behavior analysis, we utilize four sets of widely recognized public datasets, as follows.

\begin{itemize}[left=0pt]
\item \textbf{MovieLens-1M (ML-1M)\footnote{\url{https://grouplens.org/datasets/movielens/1m/}} \& MovieLens-20M (ML-20M)~\cite{harper2015movielens}\footnote{\url{https://grouplens.org/datasets/movielens/20m/}}}: These benchmark datasets in recommendation systems research contain 1 million ratings from 6,000 users on 4,000 movies (ML-1M) and 20 million ratings from 138,000 users on 27,000 movies (ML-20M). Both datasets provide user ratings from 1 to 5, along with demographic information and movie metadata.

\item \textbf{Amazon Books (AMZ-Books)~\cite{mcauley2013hidden}\footnote{\url{http://snap.stanford.edu/data/web-Amazon-links.html}}}: As a subset of the Amazon review dataset, AMZ-Books contains user reviews and ratings for books on Amazon. It offers different types of explicit feedback (e.g., ratings from 1 to 5 as well as textual reviews), making it valuable in recommendation system research.

\item \textbf{KuaiRand-27k~\cite{gao2022kuairand}\footnote{\url{http://kuairand.com}}}: A dataset from a popular short video platform that provides extensive user and content interaction data, including watching, liking, and commenting behaviors. It includes detailed user and content features such as demographics, preferences, metadata, and engagement metrics, crucial for developing and evaluating recommendation algorithms.
\end{itemize}

To evaluate the model's performance in a multi-behavior context, we also employ two prominent multi-behavior datasets:

\begin{itemize}[left=0pt]
\item \textbf{CIKM 2019 EComm AI Dataset (CIKM)\footnote{\url{https://tianchi.aliyun.com/competition/entrance/231721}}}: This dataset includes user behavior logs, product information, and interactions like clicks, add-to-cart actions, purchases, and likes. Product attributes such as category, brand, and price are also provided, along with user demographics like age, gender, and location.

\item \textbf{IJCAI-15 Repeat Buyers Prediction Dataset (IJCAI)~\cite{ijcaidata}\footnote{\url{https://tianchi.aliyun.com/dataset/42}}}: Sourced from real-world scenarios, this dataset includes user behavior records (browsing, clicking, purchasing, favoriting), product details (ID, category, brand), and basic user profiles (ID, registration info). It offers multi-dimensional information for exploring user behavior patterns.
\end{itemize}

To evaluate the performance in multi-domain recommendation, we conduct further experiments on a multi-domain dataset:
\begin{itemize}[left=0pt]
\item \textbf{Amazon Multiple Domains (AMZ-MD)} ~\cite{mcauley2013hidden}\footnote{\url{https://cseweb.ucsd.edu/~jmcauley/datasets.html\#amazon_reviews}}: Similar to AMZ-Books dataset, this dataset is selected from Amazon review dataset. For multi-domain recommendation, the interactions include users and items from four different domains: Digital Music, Movies \& TV, Toys, and Video Games.
\end{itemize}

\subsubsection{Dataset Preprocessing}
For single behavior evaluations, we use the MovieLens and Amazon Books datasets to align with evaluation  in~\cite{zhai2024actions}. For multi-behavior evaluations, we use the CIKM and IJCAI datasets from real industrial scenarios. The datasets are divided into training and test sets. In the training set, we use the whole sequence except the last two items to train and predict the second-to-last item. In the test set, we predict the last item. We follow the data preprocessing steps in~\cite{zhai2024actions}. For the multi-domain evaluations, we utilize the AMZ-MD dataset, applying a 5-core filter to exclude less popular users and items.

\subsubsection{Evaluation Metrics}

For the recall task, we adopt three widely used evaluation protocols: Hit Ratio (HR), Normalized Discounted Cumulative Gain (NDCG), and Mean Reciprocal Rank (MRR). 
HR@K is applied to measure whether the test item is under the top-K list of the recommendation results.
NDCG@K evaluates the top-K recommendation quality by giving higher scores to the
top-ranked relevant items.
MRR calculates the rank of the first relevant item presented in the recommendation results.

For the ranking task, we adopt the two most widely-used evaluation metrics: AUC and Logloss for evaluation.
AUC measures the probability of predicting higher scores of positive interactions than negative interactions. 
A higher value indicates a better performance. 
Logloss calculates the distance between the true labels and the predicted scores. 
A lower value indicates a better performance.

\subsubsection{Parameter Settings}
All experiments are implemented using PyTorch~\cite{paszke2019pytorch} on servers equipped with 8$\times$ Huawei D910B NPUs, each with 32GB of memory. We utilize the Accelerate framework\footnote{\url{https://github.com/huggingface/accelerate}} to facilitate large-scale distributed model training. To ensure a fair comparison, we maintain the original model implementations' hyper-parameters, except for those specifically being explored.

\subsection{Comparison of Model Performances (RQ1)}\label{sec:comparison-all}

We aim to investigate the impact of model depth on scaling laws by evaluating several popular transformer-based large recommendation model architectures for recall tasks in recommendation systems, including HSTU~\cite{zhai2024actions}, Llama~\cite{touvron2023llama}, GPT~\cite{achiam2023gpt}, and SASRec~\cite{kang2018selfattentive}. Our objective is to determine whether increasing the number of parameters by varying the number of attention blocks results in performance improvements and to explore further how the architecture of these backbones influences the scaling laws.

The results are detailed in Table ~\ref{tab:backbone_retrieval}. 
In our experiments, we observe that when the number of transformer blocks is low, i.e., the model has fewer parameters, the performance of the four backbones is similar across different datasets, with the best-performing architecture varying by dataset. For instance, Llama performs best on the ML-1M dataset, while GPT outperforms Llama on the AMZ-Books dataset with two blocks. When we increase the number of blocks to expand model parameters, HSTU and Llama demonstrate better scalability, while GPT and SASRec show no scalability. 
Though GPT architecture generally performs well and adheres to scaling laws in broader applications, it exhibits limited scalability in recommendation tasks. 
This could be due to the lack of architectural adaptations specific to recommendation features. 
Additionally, as shown in Table~\ref{tab:backbone_retrieval}, model performance varies with both dataset size and model parameter size, even when the architecture remains constant. To investigate this phenomenon further, we will conduct additional studies in Section~\ref{sec:ablation}.

\subsection{Understanding the Scaling Law of HSTU (RQ2)}\label{sec:ablation}

\begin{table}[t]
    \centering
    \begin{tabular}{@{}cccc@{}}
    \toprule
    \textbf{Model Variant} & \textbf{\# Blocks} & \textbf{HR@10} & \textbf{NDCG@10}\\
    \midrule
    \multirow{6}{*}{\shortstack{HSTU \\(w/o r.a.b.)}} 
    & 2  & 0.2752 & 0.1544  \\
    & 4 & 0.2944 & 0.1674 \\
    & 8 & 0.3083 &  0.1756   \\
    & 16 & 0.3135 & 0.1801  \\
    & 32 & 0.3149 & 0.1805   \\
    \midrule
    \multirow{6}{*}{\shortstack{HSTU \\(w/o SiLU)}} 
    & 2  & 0.2807 & 0.1581  \\
    & 4 & 0.3098 & 0.1761  \\
    & 8 & 0.3298 & 0.1897   \\
    & 16 & 0.3422 & 0.1992 \\
    & 32 & 0.3476  & 0.2043    \\
    \midrule
    \multirow{6}{*}{\shortstack{HSTU \\ (w/o feature interaction)}} 
    & 2  & 0.2710 & 0.1507   \\
    & 4  & 0.2900 & 0.1623  \\
    & 8 & 0.3154 & 0.1800  \\
    & 16 & 0.3300 & 0.1905 \\
    & 32 & 0.3339 & 0.1947  \\
    \bottomrule
    \end{tabular}
    \caption{Impact of various HSTU components on scaling law.}
    \label{tab:ablation-hstu-scaling}
    \vspace{-6mm}
\end{table}

Next, we investigate the origins of the scaling law in HSTU, as the representative of large recommendation models, by performing an in-depth analysis of its components. We begin with ablation studies to evaluate the impact of each key component on recommendation performance and the model's scalability. Following this, we conduct parameter analyses to assess HSTU's scalability across various hyperparameter settings. We then explore the potential to introduce a scaling law to the SASRec model by implementing specific modules informed by our ablation study findings. Finally, we examine the characteristics of HSTU that contribute to its scaling law through visualization analysis. All experiments in this section utilize the ML-20M dataset, the largest dataset in our research, with similar trends observed across other datasets.

\subsubsection{Ablation Studies} 

We investigate three key components of HSTU: the selection of relative attention bias, the use of SiLU for attention score weighting, and the method of feature interaction.

\paragraph{Impact of Components on Scaling Law}
To analyze the impact of various components of HSTU on the scaling law, we conduct an ablation study by systematically removing one key component at a time: relative attention bias (r.a.b.), the SiLU activation function, or feature interaction. We then evaluate how the model's performance varied with an increasing number of HSTU blocks. The results, presented in Table~\ref{tab:ablation-hstu-scaling}, show that most models maintain scalability even with the removal of a key component. However, the improvement in performance metrics, such as NDCG and HR, plateau most significantly when the relative attention bias is removed, particularly as the number of HSTU blocks increased from 8 to 32. This suggests that HSTU is robust, as removing a single component does not significantly impact its scalability with model depth. The qualitative results of further analyses for each individual component will be presented in the remainder of this ablation study.

\paragraph{Relative Attention Bias}
Early transformer models utilize positional embeddings to incorporate positional information between tokens effectively. However, instead of using absolute positional information, HSTU employs relative position and time difference buckets to modify attention scores. Specifically, the attention score between two items in a sequence is adjusted by a bias that depends on (1) their relative positions and (2) their time difference buckets.

We conduct a series of experiments to evaluate the impact of different attention bias mechanisms on the performance of the HSTU model. Specifically, we replace the relative attention bias module in HSTU with three alternative mechanisms: (1) Relative Attention Bias using only Bucketed Relative Time Difference (Rel. Time Diff. Bucket Only), (2) Relative Attention Bias using only Relative Position (Rel. Position Only), and (3) Rotary Positional Encoding (RoPE).
The performance results of these experiments are presented in Table~\ref{tab:ablation-rab}.

\begin{table}[t]
\centering  
    \begin{tabular}{@{}ccc@{}}
    \toprule
    \textbf{Relative Attention Bias Type} & \textbf{HR@10} & \textbf{NDCG@10} \\ 
    \midrule
    {\hspace{0.1cm}Rel. Position and Time Diff. Bucket\hspace{0.1cm}} & 0.3376 & 0.1967 \\ 
    Rel. Time Diff. Bucket Only & 0.3356 & 0.1952 \\ 
    Rel. Position Only & 0.3122 & 0.1787 \\ 
    RoPE & 0.3149  & 0.1801 \\ 
    No Attention Bias & 0.3083 & 0.1756 \\
    \bottomrule
    \end{tabular}
\caption{Ablation study on the impact of different choices of relative attention bias.}
\label{tab:ablation-rab}
\vspace{-5mm}
\end{table}
\begin{table}[t]
\centering  
    \begin{tabular}{@{} c c c @{}}
    \toprule
    \textbf{\hspace{0.1cm}Attention Score Function\hspace{0.1cm}} & \textbf{HR@10} &\textbf{NDCG@10} \\ 
    \midrule
    SiLU & 0.3376 & 0.1967 \\
    Softmax & 0.3298 & 0.1897 \\
    \bottomrule
    \end{tabular}
\caption{Ablation study on the impact of different attention score functions.}
\vspace{-5mm}
\label{tab:ablation-attention-fn}
\end{table}
\begin{table}[t]
\centering  
    \begin{tabular}{@{}ccc@{}}
    \toprule
    \textbf{Feature Interaction} & \textbf{HR@10} &\textbf{NDCG@10} \\
    \midrule
    w/ feature interaction & 0.3376 & 0.1967 \\
    {\hspace{0.1cm}w/o feature interaction} & 0.3154 & 0.1800 \\
    \bottomrule
    \end{tabular}
\caption{Ablation study on the role of feature interaction.}
\vspace{-7mm}
\label{tab:ablation-feature-interaction}
\end{table}

\begin{table*}[h!]
    \centering
    \small
    \begin{tabular}{@{} cc | >{\dope}p{1.2cm} >{\dope}p{1.2cm} >{\dope}p{1.2cm} >{\dope}p{1.2cm} >{\dope}p{1.2cm} | >{\dope}p{1.2cm} >{\dope}p{1.2cm} >{\dope}p{1.2cm} >{\dope}p{1.2cm} >{\dope}p{1.2cm} @{}}
        \toprule
          ~ & ~ & \textbf{NDCG@10} & \textbf{NDCG@50} & \textbf{HR@10} & \textbf{HR@50} & \textbf{MRR} & \textbf{NDCG@10} & \textbf{NDCG@50} & \textbf{HR@10} & \textbf{HR@50} & \textbf{MRR} \\ 
        \midrule
        \textbf{\# Dim} & \textbf{\# Blocks} & \multicolumn{5}{c}{$|S|$ = 100} & \multicolumn{5}{c}{$|S|$ = 200} \\
        \midrule
        \multirow{5}{*}{50} &12 & 0.1585 & 0.2153 & 0.2830 & 0.5407 & 0.1356 & 0.1741 & 0.2325 & 0.3083 & 0.5729 & 0.1485 \\ 
        &24 & 0.1663 & 0.2234 & 0.2937 & 0.5523 & 0.1425 & 0.1824 & 0.2408 & 0.3184 & 0.5826 & 0.1561 \\ 
        &32 & 0.1682 & 0.2253 & 0.2951 & 0.5539 & 0.1445 & 0.1867 & 0.2445 & 0.3238 & 0.5856 & 0.1599 \\ 
        &64 & 0.1684 & 0.2244 & 0.2950 & 0.5489 & 0.1445 & 0.1891 & 0.2470 & 0.3262 & 0.5881 & 0.1622 \\ 
        &96 & 0.1680 & 0.2243 & 0.2953 & 0.5473 & 0.1447 & 0.1901 & 0.2471 & 0.3277 & 0.5854 & 0.1629 \\ 
         \midrule
        \multirow{5}{*}{100}&12 & 0.1770 & 0.2337 & 0.3078 & 0.5651 & 0.1519 & 0.1952 & 0.2536 & 0.3357 & 0.6000 & 0.1674 \\ 
        &24 & 0.1805 & 0.2368 & 0.3132 & 0.5684 & 0.1547 & 0.1974 & 0.2604 & 0.3423 & 0.6022 & 0.1704 \\ 
        &32 & 0.1815 & 0.2375 & 0.3133 & 0.5669 & 0.1559 & 0.2047 & 0.2619 & 0.3493 & 0.6078 & 0.1753 \\ 
        &64 & 0.1801 & 0.2353 & 0.3118 & 0.5617 & 0.1543 & 0.2053 & 0.2619 & 0.3488 & 0.6048 & 0.1760 \\ 
        &96 & 0.1746 & 0.2296 & 0.3030 & 0.5525 & 0.1498 & 0.2032 & 0.2602 & 0.3450 & 0.6027 & 0.1746 \\ 
         \midrule
        \multirow{5}{*}{200}&12 & 0.1799 & 0.2362 & 0.3102 & 0.5649 & 0.1548 & 0.2019 & 0.2600 & 0.3441 & 0.6070 & 0.1734 \\ 
        &24 & 0.1849 & 0.2405 & 0.3171 & 0.5686 & 0.1591 & 0.2102 & 0.2673 & 0.3560 & 0.6139 & 0.1803 \\ 
        &32 & 0.1812 & 0.2365 & 0.3138 & 0.5641 & 0.1551 & 0.2120 & 0.2689 & 0.3583 & 0.6154 & 0.1819 \\ 
        &64 & 0.1752 & 0.2308 & 0.3053 & 0.5575 & 0.1500 & 0.2100 & 0.2667 & 0.3537 & 0.6098 & 0.1807 \\ 
        &96 & 0.1789 & 0.2335 & 0.3082 & 0.5555 & 0.1538 & 0.2089 & 0.2654 & 0.3532 & 0.6081 & 0.1795 \\ 
         \midrule
        \multirow{5}{*}{400}&12 & 0.1768 & 0.2327 & 0.3059 & 0.5588 & 0.1521 & 0.2037 & 0.2610 & 0.3467 & 0.6053 & 0.1748 \\ 
        &24 & 0.1767 & 0.2317 & 0.3045 & 0.5535 & 0.1521 & 0.2119 & 0.2685 & 0.3580 & 0.6132 & 0.1819 \\ 
        &32 & 0.1814 & 0.2359 & 0.3112 & 0.5577 & 0.1560 & 0.2105 & 0.2672 & 0.3555 & 0.6113 & 0.1808 \\ 
        &64 & 0.1697 & 0.2229 & 0.2935 & 0.5350 & 0.1459 & 0.2102 & 0.2670 & 0.3541 & 0.6105 & 0.1808 \\ 
        &96 & 0.1627 & 0.2205 & 0.2883 & 0.5369 & 0.1387 & 0.2033 & 0.2593 & 0.3445 & 0.5972 & 0.1747 \\ 
         \bottomrule
        
    \end{tabular}
    \caption{Parameter analyses on the ML-20M dataset with varying sequence length ($|S|$), embedding dimension (\# Dim), and number of HSTU blocks (\# Blocks).}
    \vspace{-7mm}
    \label{tab:stable1}
\end{table*}
\begin{table}[h!]
    \centering
    \begin{tabular}{@{} c c c >{\dope}p{1.5cm} >{\dope}p{1.5cm} @{}}
        \toprule
        \textbf{\hspace{0.1cm}\# Dim} & \textbf{\# Blocks} 
 &\textbf{\# Heads} & \textbf{HR@10}  & \textbf{NDCG@10}  \\ 
        \midrule
        32 & \multirow{4}{*}{8} & \multirow{4}{*}{8} & 0.2186 & 0.1156  \\ 
        64 & ~ & ~ & 0.2736  & 0.1514 \\ 
        128 & ~ & ~ & 0.2357 & 0.1770 \\ 
        256 & ~ & ~ & 0.3376 & 0.1967 \\ 
        \midrule
        \multirow{4}{*}{256} & 2 & \multirow{4}{*}{8} & 0.2928 & 0.1653 \\ 
        ~ & 4 & ~ & 0.3194 & 0.1840 \\
        ~ & 8 & ~ & 0.3376 & 0.1967	\\ 
        ~ & 16 & ~ & 0.3469 & 0.2037 \\
        \midrule
        \multirow{6}{*}{256} & \multirow{6}{*}{8} & 1 & 0.3385 & 0.1978  \\ 
        ~ & ~ & 2 & 0.3380 & 0.1968 \\
        ~ & ~ & 4 & 0.3381 & 0.1978 \\ 
        ~ & ~ & 8 & 0.3376 & 0.1967	\\ 
        ~ & ~ & 16 & 0.3380 & 0.1978 \\ 
        ~ & ~ & 32 & 0.3384 & 0.1975 \\
        \bottomrule
    \end{tabular}
    \caption{Parameter analyses of HSTU on the ML-20M dataset.}
    \vspace{-7mm}
    \label{tab:scalinglaw_ml20m}
\end{table}

Following our intuition, temporal information emerges as the most critical factor in the sequential recommendation setting, surpassing the reliance on positional information typical in language models. Our results show that models using relative attention bias with relative time information significantly outperform those using only positional information, as measured by HR and NDCG metrics. Furthermore, removing positional information from the relative attention bias does not significantly affect recall metrics. Interestingly, when using only relative positional information, RoPE enhances attention scores more effectively than relative attention bias.

\paragraph{SiLU for Attention Score Weighting}
A key innovation of HSTU, compared to other transformer-based language models, is its use of the SiLU activation function instead of softmax for calculating attention weights. We analyze how different activation functions affect attention score weighting. Specifically, we modify the spatial aggregation in the standard implementation of HSTU blocks using the following equation:
\begin{equation}
        Attn(X)V(X)=Softmax\left(\frac{Q(X)K(X)^T}{\sqrt{n}}+rab^{p,t}\right)V(X),
\end{equation}
where $X$ represents the input hidden states to the network. In our ablation study, we replace the original SiLU activation function with the Softmax function. The denominator $\sqrt{n}$ serves as the standard normalization factor in scaled dot-product attention. As shown in Table \ref{tab:ablation-attention-fn}, we observe that using the Softmax function decreases the performance of HSTU. This decline is likely because Softmax aggregation bounds the maximum attention score value, making it less expressive and therefore less suitable for recommendation tasks compared to SiLU when aggregating attention scores.

\paragraph{Feature Interaction}
As shown in Table~\ref{tab:ablation-feature-interaction}, we investigate the impact of feature-interaction mechanisms. These mechanisms are a common characteristic of deep-learning recommendation models, as demonstrated in works like DeepFM~\cite{guo2017deepfm} and DCN~\cite{wang2017deep}. These mechanisms typically involve feature-crossing, achieved through operations such as the dot product or Hadamard product between dense representations of feature pairs or dimensions within the dense representation itself. In HSTU, feature interaction is implemented via a point-wise transformation layer. To evaluate the impact of this layer, we assess HSTU's performance by removing the Hadamard product and layer normalization, modifying the point-wise transformation layer according to the following equation:
\begin{equation} 
        Y(X)=f_2(Attn(X)V(X)),
\end{equation}
where $f_2$ represents an MLP. We remove the layer normalization applied to $Attn(X)V(X)$ to enhance model stability, as another layer normalization is subsequently applied to $Y(X)$. This adjustment prevents the application of two consecutive layer normalizations without an intervening non-linear transformation. The results in Table~\ref{tab:ablation-feature-interaction} indicate that removing the Pointwise Transformation Layer significantly reduces the performance of HSTU.

\subsubsection{Parameter Analyses}
In this study, we examine scaling laws by varying embedding dimensions, the number of transformer heads, model depth, and sequence length using the ML-20M dataset. We begin by analyzing two crucial parameters: embedding dimension and the number of model layers. Table~\ref{tab:stable1} illustrates the effects of adjusting the maximum sequence length for the HSTU model on the ML-20M dataset. From this table, we derive four principal conclusions: (1) Increasing sequence length does not necessarily benefit the performance, as longer sequences may introduce more noise. (2) Performance does not consistently improve with increasing model size. Upon reaching a peak, performance begins to fluctuate, highlighting the need for guidelines to align model size with dataset size for optimal performance. (3) The optimal number of layers decreases as the embedding size increases, while the product of the optimal number of layers (L) and embedding dimension (D) remains constant, supporting our theoretical model that size is proportional to O(LD). (4) As the maximum sequence length increases, the optimal model size O(LD) also increases, indicating that larger datasets should be matched with larger models.

We also conduct an investigation into the interplay among embedding dimensions, model layers, and attention heads. The findings related to attention heads are detailed in Table~\ref{tab:scalinglaw_ml20m}. Our analysis indicates that increasing the number of HSTU blocks and the embedding dimension significantly enhances recall performance. In contrast, variations in the number of attention heads generally do not substantially affect recall performance.
\begin{table}[t]
    \centering
    \begin{tabular}{@{} cc >{\dope}p{1.2cm} >{\dope}p{1.2cm} >{\dope}p{1.2cm} @{}}
    \toprule
    \textbf{Model} & \textbf{\# Blocks} & \textbf{HR@10} & \textbf{NDCG@10} & \textbf{MRR} \\
    \midrule
    \multirow{5}{*}{SASRec} 
    & 2  & 0.2781 & 0.1553 & 0.1330  \\
    & 4 & 0.0599 & 0.0294 & 0.0284 \\
    & 8 & 0.0326 & 0.0156 & 0.0169  \\
    & 16 & 0.0349 & 0.0167 & 0.0177 \\
    & 32 & 0.0301  & 0.0159 & 0.0169   \\
    \midrule
    \multirow{5}{*}{\shortstack{SASRec\\(w/ r.a.b.)}} 
    & 2  & 0.0478 & 0.0269 & 0.0258  \\
    & 4 & 0.0447 & 0.0227 & 0.0222 \\
    & 8 & 0.0398 & 0.0213 & 0.0221  \\
    & 16 & 0.0454 & 0.0228 & 0.0222 \\
    & 32 & 0.0460  & 0.0232 & 0.0224   \\
    \midrule
    \multirow{5}{*}{\shortstack{SASRec\\(residual in HSTU)}} 
    & 2  & 0.2788 & 0.1553 & 0.1330 \\
    & 4 & 0.2551 & 0.1398 & 0.1199 \\
    & 8 & 0.2059 & 0.1085 & 0.0939  \\
    & 16 & 0.0628 & 0.0297 & 0.0296 \\
    & 32 & 0.0589  & 0.0282 & 0.0282   \\
    \midrule
    \multirow{5}{*}{\shortstack{SASRec\\(residual in Llama)}} 
    & 2  & 0.2850 & 0.1592 & 0.1359 \\
    & 4 & 0.2575 & 0.1422 & 0.1223 \\
    & 8 & 0.1380 & 0.0699 & 0.0633  \\
    & 16 & 0.1242 & 0.0617 & 0.0562 \\
    & 32 & 0.1234  & 0.0609 & 0.0552   \\
    \midrule
    \multirow{5}{*}{\shortstack{SASRec\\
    (residual in Llama, \\w/ r.a.b.)}} 
    & 2  & 0.2990 & 0.1689 & 0.1445 \\
    & 4 & 0.3116 & 0.1799 & 0.1513 \\
    & 8 & 0.3152 & 0.1809 & 0.1571  \\
    & 16 & 0.3140 & 0.1796 & 0.1538 \\
    & 32 & 0.3182  & 0.1835 & 0.1575   \\
    \bottomrule
    \end{tabular}
    \caption{Analyses on the scaling law of SASRec on the ML-20M dataset.}
    \vspace{-5mm}
    \label{tab:ablation-sasrec}
\end{table}

\begin{figure}[t]
    \centering
    \includegraphics[width=\linewidth]{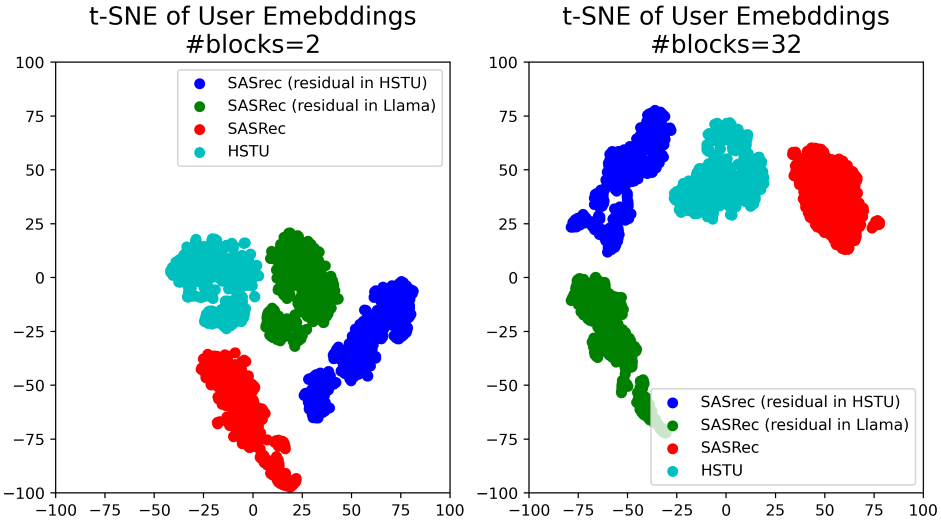}
    \caption{Visualization of user embeddings on the ML-20M dataset.}
    \label{fig:visual}
        \vspace{-5mm}
\end{figure}

\subsubsection{Scaling Law in SASRec}
To further investigate the source of the scaling law, we examine whether the success of HSTU and Llama can be replicated by modifying the self-attention layers in traditional backbone models like SASRec. Specifically, we introduce the relative attention bias (r.a.b.) from HSTU into the standard self-attention layers of SASRec by incorporating it into the $Q(X)K(X)^{T}$ computation during attention processing. Additionally, we adapt the residual connection using the patterns implemented in HSTU and Llama, as described below:

\begin{align}
&\text{HSTU}\hspace{0.2cm}\left\{\hspace{0.2cm}\label{eqn:residual-hstu}
\begin{aligned}
    & g_{1}(X) = SA(LN(X)),&\\
    & g_{2}(X) = FFN(LN(X)),& \\
    & B_{l}(X) = X + g_{2}(g_{1}(X)),&
\end{aligned}
\right.
&\\
& \nonumber\\
    &\text{Llama}\hspace{0.2cm}\left\{\hspace{0.2cm}\label{eqn:residual-llama}
\begin{aligned}
     &g_{1}(X) = X + SA(LN(X)),&\\
     &g_{2}(X) = X + FFN(LN(X)),& \\
     &B_{l}(X) = g_{2}(g_{1}(X)),&
\end{aligned}
\right.
\end{align}
where $B_{l}(X)$ denotes the output of the $l$-th attention block, given the hidden state $X$ from the $(l-1)$-th block. The components of the attention blocks include $LN(\cdot)$ for layer normalization, $SA(\cdot)$ for self-attention layers, and $FFN(\cdot)$ for feed-forward networks. A key distinction in our approach is the implementation of the residual connection \textit{before} the layer normalization operation, in contrast to the SASRec framework, which applies it \textit{after} layer normalization. The results of our experiments using the ML-20M dataset are presented in Table~\ref{tab:ablation-sasrec}.

Our experiments reveal that increasing the number of transformer blocks degrades the performance of SASRec on the ML-20M dataset, thereby highlighting the poor scalability of the original SASRec model. This negative trend persists regardless of whether we incorporate the relative attention bias (r.a.b.) module or modify the residual connection as described in Equations~\ref{eqn:residual-hstu} and \ref{eqn:residual-llama}. Notably, the residual connection pattern used in Llama exhibits greater robustness to an increasing number of blocks. Consequently, we apply the r.a.b. module to the residual connections in Llama, which demonstrates improved scalability on the ML-20M dataset. These findings suggest that both the residual connection pattern and the relative attention bias contribute to enhancing the scalability of traditional recommendation models. These preliminary evaluations provide valuable insights for future research on the scaling laws of recommendation systems.

\subsubsection{Visualization Analysis} 

\begin{table*}[t]
    \centering
    \begin{tabular}{@{} c c c >{\dope}p{1.3cm} >{\dope}p{1.3cm} >{\dope}p{1.3cm} >{\dope}p{1.3cm} >{\dope}p{1.3cm} @{}}
        \toprule
        \textbf{Dataset} & \textbf{Side Info.} & \textbf{\# Blocks} 
  & \textbf{HR@10} & \textbf{HR@50} & \textbf{NDCG@10} & \textbf{NDCG@50} & \textbf{MRR} \\ 
        \midrule
        \multirow{10}{*}{ML-1M} & \multirow{5}{*}{w/o} & 2 & 0.2962 & 0.5757 & 0.1662 & 0.2283 & 0.1427  \\ 
        ~ & ~ & 4  & 0.3206 & 0.5856 & 0.1791 & 0.2379 & 0.1513 \\ 
        ~ & ~ & 8  & 0.3312 & 0.5991 & 0.1871 & 0.2468 & 0.1588 \\ 
        ~ & ~ & 16  & 0.3333 & 0.5973 & 0.1881 & 0.2468 & 0.1590  \\ 
        ~ & ~ & 32  & 0.3289 & 0.5946 & 0.1842 & 0.2431 & 0.1554  \\ 
        \cmidrule(l){2-8}
        ~ & \multirow{5}{*}{w/} & 2 & 0.3011 & 0.5768 & 0.1693 & 0.2305 & 0.1450  \\ 
        ~ & ~ & 4 & 0.3177 & 0.5908 & 0.1795 & 0.2402 & 0.1531 \\
        ~ & ~ & 8 & 0.3236 & 0.5998 & 0.1870 & 0.2485 & 0.1613  \\ 
        ~ & ~ & 16 & 0.3242 & 0.5975 & 0.1841 & 0.2449 & 0.1569 \\
        ~ & ~ & 32 & 0.3345 & 0.6003 & 0.1881 & 0.2472 & 0.1587 \\
        \midrule
        \multirow{10}{*}{ML-20M} & \multirow{5}{*}{w/o} & 2 & 0.2907 & 0.5560 & 0.1639 & 0.2225 & 0.1407  \\ 
        ~ & ~ & 4 & 0.3176 & 0.5818 & 0.1817 & 0.2402 & 0.1556 \\
        ~ & ~ & 8 & 0.3407 & 0.6009 & 0.1991	& 0.2567 & 	0.1708  \\ 
        ~ & ~ & 16 & 0.3517 & 0.6116 & 0.2076 & 0.2651 & 0.1783 \\
        ~ & ~ & 32 & 0.3597 & 0.6148 & 0.2126 & 0.2691 & 0.1821 \\
        \cmidrule(l){2-8}
        ~ & \multirow{5}{*}{w/} & 2 & 0.2877 & 0.4428 & 0.1622 & 0.2207 & 0.1394  \\ 
        ~ & ~ & 4 & 0.3162 & 0.5797 & 0.1812 & 0.2394 & 0.1552 \\
        ~ & ~ & 8 & 0.3355 & 0.5971 & 0.1952	& 0.2531 & 	0.1674  \\ 
        ~ & ~ & 16 & 0.3493 & 0.6094 & 0.2048 & 0.2262 & 0.1755 \\
        ~ & ~ & 32 & 0.3575 & 0.6136 & 0.2110 & 0.2677 & 0.1808 \\
        \bottomrule
    \end{tabular}
    \caption{Comparison of the HSTU model's performance with and without side information. ``w/'' indicates the inclusion of side information, while ``w/o'' denotes its absence.}
    \label{tab:side_info_ml1m_ml20m}
    \vspace{-5mm}
\end{table*}
Previous analyses have indicated that components such as residual connections and relative attention bias can influence the scalability of recommendation models. Additionally, factors like dimension size and sequence length affect recall performance. However, the specific model characteristics that enhance scalability remain unclear. To investigate the intrinsic factors influencing model scalability, we visualize user embeddings generated by four model variants: HSTU, SASRec, SASRec (w/ residual in HSTU, cf. Equation~\ref{eqn:residual-hstu}), and SASRec (w/ residual in Llama, cf. Equation~\ref{eqn:residual-llama}). We randomly sample 1,000 users from the ML-20M dataset and evaluate the embeddings with 2 and 32 transformer blocks, resulting in a total of 8,000 samples. We apply t-SNE to visualize these embeddings, with the results presented in Figure~\ref{fig:visual}.

The visualization in Figure~\ref{fig:visual} reveals that embeddings in the shallower model (with 2 blocks) are more clustered compared to those in the deeper model (with 32 blocks). Notably, the HSTU model's embeddings are closest to the coordinate (0,0) in both settings. Given that HSTU demonstrates superior performance in accuracy and scalability, as shown in Tables~\ref{tab:backbone_retrieval} and~\ref{tab:ablation-sasrec}, we infer that a well-normalized model may enhance both scalability and recall performance. This suggests that normalization is crucial in enhancing the effectiveness of both shallower and deeper models.

\subsection{Evaluating HSTU in Complex User Behavioral Sequence Modeling (RQ3)}\label{sec:complex-sequence-modeling}

In this subsection, we further analyze the capacity of large recommendation models in complex user behavioral sequences. We use HSTU as the representative model for the evaluation. Specifically, we concentrate on three primary scenarios: behavior modeling with side information, multi-behavior sequence modeling, and multi-domain sequence modeling.

\subsubsection{Behavior Modeling with Side Information}

In this section, we investigate the effect of incorporating side information on the user behavior modeling process. We evaluate the performance of HSTU using two datasets, ML-1M and ML-20M, comparing scenarios with and without item attributes. The ML-1M and ML-20M datasets represent the smallest and largest data sizes in our study, respectively, providing a comprehensive range for analysis. These attributes, including movie metadata, are converted into dense vectors and combined with item ID-based embeddings using mean pooling. Additionally, we vary the model depth to investigate how the inclusion of side information influences the scaling behavior of large recommendation models.

The results are detailed in Table~\ref{tab:side_info_ml1m_ml20m}.
First, we observe that incorporating side information does not necessarily enhance the model's performance. In most instances, the inclusion of side information leads to slightly inferior results compared to the model without it. We hypothesize that this unexpected outcome may be attributed to several factors: (1) the side information employed is relatively simplistic and does not provide significant benefits to the user behavior modeling process, and (2) the method of integrating side information with item ID-based embeddings through mean pooling may be insufficient for extracting meaningful insights from the movie metadata. Future work could explore datasets with more comprehensive side information and employ more sophisticated techniques for integrating side information with user behavior sequences.

Next, we observe that the model maintains its scalability even with the incorporation of side information. This is a desirable attribute, as deeper networks are typically more challenging to train. As shown in Table~\ref{tab:side_info_ml1m_ml20m}, the model's performance consistently improves with increased depth, using up to 32 blocks. These results indicate that the integration of side information in large recommendation models remains a promising area for further research.

\subsubsection{Multi-behavior Modeling}

\begin{table*}[t]
    \centering
    \resizebox{\textwidth}{!}{
        \begin{tabular}{@{} cc cc cc cc cc cc @{}}
            \toprule
            \textbf{Dataset} & \textbf{Model} & \textbf{Training} & \textbf{Test} & \textbf{HR@50} & \textbf{NDCG@50} &
            \textbf{Dataset} & \textbf{Model} & \textbf{Training} & \textbf{Test} & \textbf{HR@50} & \textbf{NDCG@50} \\ 
            \midrule
            \multirow{12}{*}{CIKM} & \multirow{2}{*}{SASRec} & \multirow{2}{*}{all} & buy & 0.1428 & 0.0491 & \multirow{12}{*}{IJCAI} & \multirow{2}{*}{SASRec} & \multirow{2}{*}{all} & buy & 0.1207 & 0.0488 \\
                 &       &     & pv  & 0.1653 & 0.0642 &       &       &     & pv  & 0.1729 & 0.0687 \\
            \cmidrule(l){2-6} \cmidrule(l){8-12}
            ~ & \multirow{2}{*}{MBSTR} & \multirow{2}{*}{all} & buy & 0.0922 & 0.0317 & ~ & \multirow{2}{*}{MBSTR} & \multirow{2}{*}{all} & buy & 0.0960 & 0.0339 \\
              &       &     & pv  & 0.2208 & 0.0819 &   &       &     & pv  & 0.2307 & 0.0863 \\
            \cmidrule(l){2-6} \cmidrule(l){8-12}
            ~ & \multirow{8}{*}{HSTU} & \multirow{2}{*}{buy} & buy & 0.0422 & 0.0177 & ~ & \multirow{8}{*}{HSTU} & \multirow{2}{*}{buy} & buy & 0.0596 & 0.0222 \\
              &       &     & pv  & 0.0264 & 0.0148 &   &       &     & pv  & 0.0449 & 0.0172 \\
            \cmidrule(l){3-6} \cmidrule(l){9-12}
            ~ &       & \multirow{2}{*}{pv}  & buy & 0.1104 & 0.0452 & ~ &       & \multirow{2}{*}{pv}  & buy & 0.1108 & 0.0386 \\
              &       &     & pv  & 0.1089 & 0.0433 &   &       &     & pv  & 0.2133 & 0.0823 \\
            \cmidrule(l){3-6} \cmidrule(l){9-12}
            ~ &       & \multirow{2}{*}{buy \& pv} & buy & 0.1455 & 0.0575 & ~ &       & \multirow{2}{*}{buy \& pv} & buy & 0.1060 & 0.0365 \\
              &       &           & pv  & 0.1510 & 0.0609 &   &       &           & pv  & 0.2255 & 0.0848 \\
            \cmidrule(l){3-6} \cmidrule(l){9-12}
            ~ &       & \multirow{2}{*}{all} & buy & 0.1431 & 0.0558 & ~ &       & \multirow{2}{*}{all} & buy & 0.1049 & 0.0363 \\
              &       &     & pv  & 0.1761 & 0.0696 &   &       &     & pv  & 0.2301 & 0.0875 \\
            \bottomrule
        \end{tabular}
    }
    \caption{Performance comparison on multi-behavior sequences. The “all” in the training set represents user sequences that contain all four behaviors.}
    \label{tab:table1_1}
    \vspace{-7mm}
\end{table*}

\begin{table}[t]
    \centering
    \begin{tabular}{@{}lccccc@{}}
    \toprule
    \textbf{Model} & \textbf{Training} & \textbf{Test} & \textbf{HR@50} & \textbf{NDCG@50} \\
    \midrule
    \multirow{2}{*}{\centering HSTU} & \multirow{2}{*}{\centering all} & buy & 0.1431 & 0.0558 \\
          &  & pv  & 0.1761 & 0.0696 \\
    \multirow{2}{*}{\centering HSTU (w/b)} & \multirow{2}{*}{\centering all} & buy & 0.1463 & 0.0566 \\
          &  & pv  & 0.1780 & 0.0712 \\
    \bottomrule
    \end{tabular}
    \caption{Performance of the HSTU trained on the CIKM dataset. The ``w/ b'' variant denotes the inclusion of explicit behavior modeling. ``all'' in the training set column refers to user sequences encompassing all four specified behaviors.}
    \vspace{-8mm}
    \label{tab:table2_2}
\end{table}

We evaluate the HSTU model for multi-behavior modeling through three experiments: (1) the impact of using multiple behavior training data, (2) the impact of explicit behavior modeling, and (3) the exploration of scaling law in multi-behavior scenarios. For this evaluation, we use two multi-behavior datasets, CIKM and IJCAI, to construct user behavior sequences.

\paragraph{Impact of Multiple Behavior Training Data.}
Users' multi-behavior sequences are constructed by combining different types of user behaviors in chronological order, which we refer to as ``all''. To thoroughly evaluate the impact of multi-behavior data, we create distinct subsets by selectively choosing specific behaviors and combining them in chronological order. Specifically, the training set ``buy'' includes only purchase behaviors, ``pv'' comprises solely page view behaviors, and ``buy \& pv'' is a chronological concatenation of purchase and page view behaviors. The model's performance is assessed with page views and purchases as target behaviors. To comprehensively evaluate the effectiveness of the HSTU model, we compare it against several state-of-the-art multi-behavior recommendation models, including SASRec~\cite{kang2018selfattentive} and MBSTR~\cite{yuan2022multibehavior}. For a fair comparison, all models are configured with identical parameter settings: four transformer blocks, four attention heads, a batch size of 512, and a learning rate of 1e-3.

The experimental results are shown in Table~\ref{tab:table1_1}. Overall, the HSTU model outperforms the baseline models (SASRec and MBSTR), though it shows slightly worse performance on certain partitioned datasets. This suggests that while generative models like HSTU are generally superior to traditional models, they may be sensitive to specific data distributions. Incorporating a greater variety of behavioral data into the training set tends to improve recall performance on the target behavior.  For instance, in the CIKM dataset, training with page view data (``pv'') yields an HR@50 value of 0.1089 for page views. Adding purchase data (``buy \& pv'') increases the HR@50 to 0.1510, and training with all behavior types further increases it to 0.1761. This demonstrates that exposure to a broader range of behavioral data during training significantly enhances model performance.
This finding suggests that recommendation models can benefit from enriched training data, similar to large language models (LLMs). Future work should explore the integration of additional behavior types and multi-modal recommendation data to further improve model efficacy.

The results for purchase behavior presented in Table~\ref{tab:table1_1} further reinforce this observation. Training on purchase behavior alone yields an HR@50 of 0.0422, which is significantly lower than training on page view behavior, which achieves an HR@50 of 0.1104. This occurs even though the purchase data aligns with the test set distribution, suggesting that larger training sets, even with slightly lower data quality, can still substantially improve performance. However, increasing dataset size does not always lead to better performance. For instance, when comparing training on the full CIKM dataset (``all'') versus just purchase and page view behaviors (``buy \& pv''), the results for purchase behavior are similar, with some metrics declining slightly. This may be due to the limited correlation and smaller volume of additional behaviors (e.g., add to cart, like) in relation to purchase behavior. Consequently, when the increase in data size is marginal and the quality of the additional data is lower, it can have an adverse impact on overall performance.
Future work will focus on thoroughly evaluating the correlation and dependencies between different types of user behaviors and constructing high-quality multi-behavior datasets that effectively strike a balance between data quantity and quality to further enhance model performance and robustness.

\begin{figure}[t]
    \centering
    \begin{subfigure}{0.231\textwidth}
        \centering
        \includegraphics[width=\linewidth]{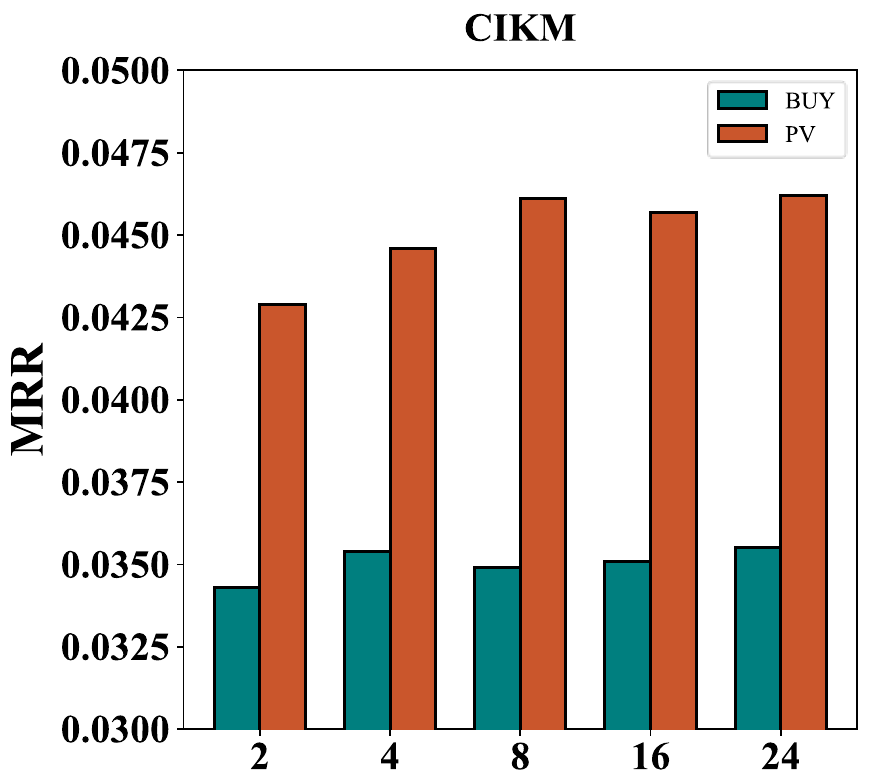}
        \caption{HSTU}
        \label{fig:fig1_1_without_b}
    \end{subfigure}
    \hfill
    \begin{subfigure}{0.231\textwidth}
        \centering
        \includegraphics[width=\linewidth]{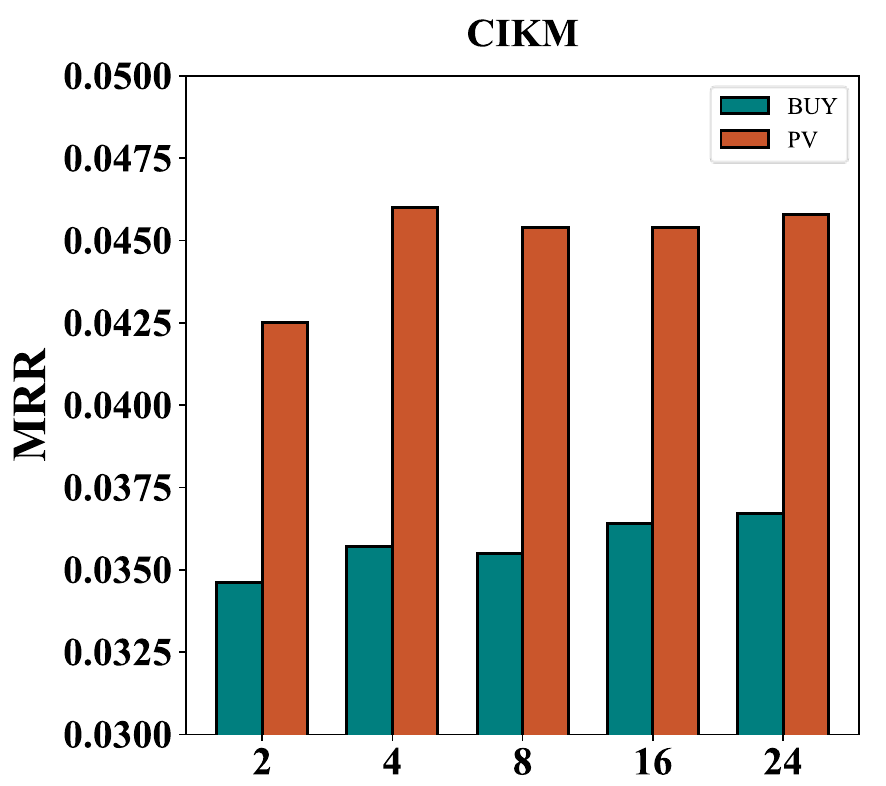}
        \caption{HSTU w/b}
        \label{fig:fig1_1_with_b}
    \end{subfigure}
    \vspace{-3mm}
    \caption{Comparison of the scalability performance between the HSTU model and the HSTU w/b variant on the CIKM dataset. The figure illustrates how each model scales with an increasing transformer layer.}
    \label{fig:fig1_1}
 \end{figure}

\begin{table*}[t]
\begin{tabular}{c >{\dope}p{1.5cm} >{\dope}p{1.5cm} >{\dope}p{1.5cm} >{\dope}p{1.5cm} >{\dope}p{1.5cm} >{\dope}p{1.5cm} >{\dope}p{1.5cm} >{\dope}p{1.5cm}}
    \toprule
    \multirow{2}{*}{\textbf{Model}} & \multicolumn{2}{c}{\textbf{Digital Music}} & \multicolumn{2}{c}{\textbf{Movies \& TV}} & \multicolumn{2}{c}{\textbf{Toys}} & \multicolumn{2}{c}{\textbf{Video Games}} \\ 
    ~ & \textbf{HR@10} & \textbf{NDCG@10} & \textbf{HR@10} & \textbf{NDCG@10} & \textbf{HR@10} & \textbf{NDCG@10} & \textbf{HR@10} & \textbf{NDCG@10} \\
    \midrule
    SASRec & 0.1332 & 0.0764 & 0.0977 & 0.0577 & 0.0758 & 0.0441 & 0.1228 & 0.0677 \\ 
    C2DSR & 0.1363 & 0.0772 & 0.0974 & 0.0578 & 0.0745 & 0.0442 & 0.1194 & 0.0658 \\ 
    HSTU & 0.1451 & 0.0860 & 0.1020 & 0.0598 & 0.0704 & 0.0425 & 0.1330 & 0.0738 \\ 
    HSTU-single & 0.1004 & 0.0577 & 0.1022 & 0.0597 & 0.0588 & 0.0337 & 0.1055 & 0.0576\\
    \bottomrule
\end{tabular}
\caption{Performance comparison for cross-domain recommendation on different target datasets. The ``HSTU-single'' model variant is trained with a single domain dataset.}
\vspace{-4mm}
\label{tab:MDR_performance}
\end{table*}
\begin{table*}[t]
\begin{tabular}{c >{\dope}p{1.5cm} >{\dope}p{1.5cm} >{\dope}p{1.5cm} >{\dope}p{1.5cm} >{\dope}p{1.5cm} >{\dope}p{1.5cm} >{\dope}p{1.5cm} >{\dope}p{1.5cm}}
    \toprule
    \multirow{2}{*}{\textbf{\# Blocks}} & \multicolumn{2}{c}{\textbf{Digital Music}} & \multicolumn{2}{c}{\textbf{Movies \& TV}} & \multicolumn{2}{c}{\textbf{Toys}} & \multicolumn{2}{c}{\textbf{Video Games}} \\
    ~ & \textbf{HR@10} & \textbf{NDCG@10} & \textbf{HR@10} & \textbf{NDCG@10} & \textbf{HR@10} & \textbf{NDCG@10} & \textbf{HR@10} & \textbf{NDCG@10} \\
    \midrule
    2 & 0.1225 & 0.0704 & 0.0909 & 0.0532 & 0.0645 & 0.0383 & 0.1166 & 0.0643 \\
    4 & 0.1338 & 0.0773 & 0.0957 & 0.0564 & 0.0669 & 0.0399 & 0.1237 & 0.0686 \\
    8 & 0.1353 & 0.0805 & 0.0971 & 0.0565 & 0.0671 & 0.0404 & 0.1235 & 0.0672 \\
    16 & 0.1489 & 0.0874 & 0.1031 & 0.0596 & 0.0685 & 0.0404 & 0.1365 & 0.0764 \\
    32 & 0.1556 & 0.0937 & 0.1090 & 0.0625 & 0.0716 & 0.0416 & 0.1422 & 0.0781 \\
    \bottomrule
\end{tabular}
\caption{Performance comparison for cross-domain recommendation on different target datasets with varying number of HSTU blocks.}
\vspace{-5mm}
\label{tab:MDR_HSTUlayers}
\end{table*}

\paragraph{Impact of Explicit Behavior Modeling.} In previous research, models often treated behaviors within sequences as indistinguishable. For instance, when purchase  and page view behaviors were concatenated chronologically, the model could not distinguish between them in the input sequence. In this section, we explore whether explicitly modeling behavior types can enhance performance. We achieve this by representing each behavior explicitly as a token in the input sequence, formatted as \(X_u = \{x_1, b_1, x_2, b_2, \ldots, x_n, b_n\}\). Using the same training set, we conduct experiments across two subsets: single behavior (``buy'' or ``pv'' only). 
The results in Table~\ref{tab:table2_2} indicate that the model incorporating explicit behavior modeling (``HSTU w/ b'') generally performs better than the model without this feature (``HSTU''). This demonstrates that adding explicit behavior tokens greatly enhances the model’s ability to capture user interaction nuances, leading to improved performance.

\paragraph{Scaling Law in Multi-Behavior Scenarios.} To investigate the scaling law of HSTU in multi-behavior scenarios, we systematically increase the number of model layers from 2 to 24. As illustrated in Figures~\ref{fig:fig1_1}, we observe that, with the increase in the number of layers, the overall performance generally improves, with a few exceptions. This trend holds consistently across the IJCAI dataset in our experiments, highlighting the benefits of scaling up model parameters and demonstrating the potential advantages of larger recommendation models.



\subsubsection{Multi-domain Modeling} 
In this section, we evaluate the performance of HSTU in multi-domain modeling. We select several state-of-the-art multi-domain recommendation models as baselines, including C2DSR~\cite{Cao2022contrastive} and SASRec, to ensure a comprehensive comparison. HSTU serves as the backbone model to test its performance and scalability in multi-domain recommendation contexts. We conduct experiments on a multi-domain dataset AMZ-MD.

\paragraph{Performance of HSTU for Multi-domain Recommendation.} 
In practical applications, user behavior data is often collected from multiple scenarios, enabling large recommendation models to learn across domains. We evaluate HSTU's performance on the AMZ-MD dataset and compare it with the baseline models. As shown in Table~\ref{tab:MDR_performance}, HSTU outperforms other recommendation models across most domains, though it exhibits some performance gaps in the Toys domain. This suggests large recommendation models can more effectively learn from multi-domain interaction data due to their extensive parameter capacity. However, when trained on single-domain datasets, HSTU's performance significantly declines in all domains except Movies \& TV, highlighting its strength and potential in multi-domain scenarios. The results in the Movies \& TV domain indicate that this domain is slightly influenced by knowledge transfer from other domains, with performance primarily determined by the HSTU's single-domain modeling capability.

\paragraph{Scaling Law in HSTU for Multi-domain Recommendation.} To explore the scaling law in HSTU, we conduct experiments by varying the number of layers in multi-domain datasets. The results, shown in Table~\ref{tab:MDR_HSTUlayers}, indicate that as the model's complexity increases, performance improves across all domains, paralleling trends observed in single-domain experiments. This improvement is particularly notable in domains with fewer items and interactions, such as Digital Music and Video Games, suggesting that larger models facilitate enhanced cross-domain knowledge transfer. This finding implies that HSTU could be instrumental in addressing the cold-start problem.

\begin{table*}[h!]
    \centering
    \begin{tabular}{@{} c c >{\dope}p{1.3cm} >{\dope}p{1.3cm} >{\dope}p{1.3cm} >{\dope}p{1.3cm} >{\dope}p{1.3cm} >{\dope}p{1.3cm} @{}}
    \toprule
    \multirow{2}{*}{\textbf{Model}} & \multirow{2}{*}{\textbf{\# Blocks}} & \multicolumn{2}{c}{\textbf{ML-1M}} & \multicolumn{2}{c}{\textbf{ML-20M}} & \multicolumn{2}{c}{\textbf{AMZ-Books}} \\
    & & \textbf{AUC} & \textbf{Logloss} & \textbf{AUC} & \textbf{Logloss} & \textbf{AUC} & \textbf{Logloss} \\
    \midrule
    \multirow{1}{*}{DIN}
    & - & 0.7241 & 0.6141 & 0.7247 & 0.6135 & 0.7060 & 0.4562 \\
    \midrule
    \multirow{6}{*}{HSTU} 
    & 2  & 0.7559 & 0.5814 & 0.7813 & 0.5539 & 0.7257 & 0.4608 \\
    & 4  & 0.7530 & 0.5821 & 0.7920 & 0.5422 & 0.7386 & 0.4682 \\
    & 8  & 0.7591 & 0.5772 & 0.7960 & 0.5394 & 0.7283 & 0.5134 \\
    & 16 & 0.7943 & 0.5318 & 0.7879 & 0.5463 & 0.7442 & 0.5089 \\
    & 24 & 0.7943 & 0.5307 & 0.7992 & 0.5360 & 0.7450 & 0.4496 \\
    & 32 & 0.7947 & 0.5341 & 0.7914 & 0.5416 & 0.7606 & 0.4140 \\
    \midrule
    \multirow{6}{*}{\hspace{0.5cm}Llama\hspace{0.5cm}} 
    & 2  & 0.7922 & 0.5403 & 0.7568 & 0.5878 & 0.7181 & 0.5175 \\
    & 4  & 0.7923 & 0.5592 & 0.7595 & 0.5732 & 0.7585 & 0.4183 \\
    & 8  & 0.7939 & 0.5454 & 0.7375 & 0.5940 & 0.7449 & 0.4868 \\
    & 16 & 0.7915 & 0.5422 & 0.6390 & 0.6790 & 0.7469 & 0.4896 \\
    & 24 & 0.7883 & 0.5495 & 0.5777 & 0.6738 & 0.7517 & 0.4250 \\
    & 32 & 0.7923 & 0.5453 & 0.7107 & 0.6127 & 0.7491 & 0.4653 \\
    \bottomrule
    \end{tabular}
    \caption{Performance comparison for the ranking task. A higher value indicates a better performance for the AUC metric, while a lower value indicates a better performance for the Logloss metric.}
    \vspace{-7mm}
    \label{tab:ranking}
\end{table*}

\subsection{Evaluating HSTU in Ranking Tasks (RQ4)}\label{sec:ranking-task}

In this subsection, we aim to explore whether HSTU is still effective and scalable in ranking tasks.
\subsubsection{Experimental Settings}

In the ranking task, the input sequence \(X_u = \{x_1, b_1', x_2, b_2', \ldots, x_n, b_n'\}\) is processed to produce an output sequence. Formally, this is expressed as \(Transformer(X_u) = \{b_1', x_2, b_2', x_3, \ldots, b_n', x_{n+1}\}\). 
At each position in the sequence, a shared small neural network is attached to predict the label \(b_i'\) of the user clicking on the item \(x_i\). 

To optimize the ranking model's parameters, we employ binary cross-entropy loss \(\mathcal{L}_{ranking} = \sum_{u \in \mathcal{U}} \sum_{k=1}^{n} \text{BCELoss}(b_k'', b_k')\) to minimize the differences between prediction $b_k''$ and ground-truth label $b_k'$. We validate the effectiveness of this ranking model through experiments conducted on the ML-1M, ML-20M, and AMZ-Books datasets. We transform user-item interaction in these datasets into a binary feedback format. Specifically, interactions where users rate items with a score of 4 or 5 are categorized as positive feedback and assigned a value of 1. Conversely, interactions with ratings below 4 are considered non-positive feedback and assigned a value of 0. 

\subsubsection{Ranking Performance}
Table~\ref{tab:ranking} highlights the superior performance of HSTU compared to Llama as well as the traditional recommendation model DIN. 
Within the generative framework, Llama generally demonstrates better performance than HSTU when both have the same relatively small number of blocks. However, upon expanding the transformer blocks, we observe that the performance improvement with Llama is minimal and, in some cases, slightly decreases. In contrast, HSTU exhibits better scalability, generally improving as the number of blocks increases.

Our experiments indicate that discrepancies can occur between Logloss and AUC metrics. For instance, although the HSTU model with 32 blocks achieves a higher AUC on the ML-1M dataset compared to the HSTU model with 24 blocks, its Logloss performance is inferior. This finding highlights that a reduction in the Logloss does not consistently correspond to an improvement in AUC, suggesting that a decrease in loss does not necessarily equate to enhanced performance in the recommendation domain. This observation implies that focusing solely on scaling laws may be insufficient; it is essential to investigate underlying performance laws to gain a more comprehensive understanding.

\subsubsection{Impact of the Number of Negative Samples}
\begin{table}[t]
    \centering
    \begin{tabular}{@{} cc ccc @{}}
    \toprule
    \multirow{2}{*}{\textbf{Model}} & \multirow{2}{*}{\textbf{\shortstack{Sampling\\Ratio}}} & \multicolumn{3}{c}{\textbf{AUC}} \\
    ~ & ~ & \textbf{ML-1M} & \textbf{ML-20M} & \textbf{AMZ-Books} \\
    \midrule
    \multirow{5}{*}{HSTU} 
    & 0.2 & 0.7468 & 0.7903 & 0.6925 \\
    & 0.4 & 0.7399 & 0.7952 & 0.7093 \\
    & 0.6 & 0.7626 & 0.7950 & 0.7324 \\
    & 0.8 & 0.7622 & 0.7899 & 0.7338 \\
    & 1.0 & 0.7794 & 0.7916 & 0.7037 \\
    \midrule
    \multirow{5}{*}{Llama} 
    & 0.2 & 0.7866 & 0.7357 & 0.6824 \\
    & 0.4 & 0.7939 & 0.7376 & 0.7159 \\
    & 0.6 & 0.7930 & 0.7260 & 0.7313 \\
    & 0.8 & 0.7916 & 0.7394 & 0.7414 \\
    & 1.0 & 0.7927 & 0.7495 & 0.7433 \\
    \bottomrule
    \end{tabular}
    \caption{Performance comparison for the ranking task under varying negative sampling ratios.}
    \vspace{-5mm}
    \label{tab:Negative_Sampling_Ratios}
\end{table}

In real-world applications, datasets often exhibit an imbalanced ratio of positive to negative samples, with significantly fewer positive samples compared to negative ones. This imbalance can skew the gradient updates during model training, hindering the model's ability to accurately learn user behavior patterns. To mitigate the challenges posed by this imbalance, traditional approaches selectively sample a subset of negative samples from those the user has interacted with, thereby controlling the proportion of negative samples. 

While sampling negative samples can improve the performance of recommendation models, it is not an optimal solution, as it primarily addresses the models' limited capacities to handle complex data. Specifically, the imbalance between positive and negative samples often results in low-quality data, posing a significant challenge for traditional models. These models typically simplify the problem by randomly sampling negative instances, which can lead to the loss of critical information. In this section, we examine whether generative recommendation models can better handle such complexities. We specifically investigate how varying the ratios of negative samples affects model performance.

We randomly sample negative samples at ratios from the set \{0.2, 0.4, 0.6, 0.8, 1.0\}, where a ratio of 1.0 reflects the model's performance on the original dataset.  
The experimental results are shown in Table~\ref{tab:Negative_Sampling_Ratios}. It can be observed that increasing the sampling ratio and thereby augmenting the number of negative samples, leads to a continuous improvement in the model's performance. This improvement suggests that the generative recommendation model is adept at handling complex data and possesses superior modeling capabilities, as it benefits from the enriched information provided by the additional negative samples. Consequently, it is well-suited for complex real-world scenarios, offering significant application potential. Furthermore, the performance improvements are more pronounced in the larger dataset (ML-20M) compared to the smaller dataset (ML-1M). This observation underscores the importance of understanding data scaling laws: while expanding the dataset generally enhances model performance, the benefits appear to exhibit diminishing returns as the dataset size increases.
\begin{table}[t]
    \centering
    \begin{tabular}{@{}ccccc@{}}
    \toprule
    \multirow{2}{*}{\textbf{Model}} & \multirow{2}{*}{\textbf{Architecture}} & \multicolumn{3}{c}{\textbf{AUC}} \\
    & & \textbf{ML-1M} & \textbf{ML-20M} & \textbf{AMZ-Books} \\
    \midrule
    \multirow{3}{*}{HSTU} 
    & Dot         & 0.7886 & 0.7576 & 0.7468 \\
    & MLP         & 0.7489 & 0.7913 & 0.7386 \\
    & FFN & 0.7765 & 0.7920 & 0.7348 \\
    \midrule
    \multirow{3}{*}{Llama} 
    & Dot         & 0.7945 & 0.7137 & 0.7177 \\
    & MLP         & 0.7923 & 0.7311 & 0.7585 \\
    & FFN & 0.7901 & 0.7573 & 0.7402 \\
    \bottomrule
    \end{tabular}
    \caption{Performance comparison for the ranking task with different scoring network architectures.}
    \vspace{-6mm}
    \label{tab:ranking_architecture}
\end{table}

\begin{table}[t ]
    \centering
    \begin{tabular}{@{}ccccc@{}}
    \toprule
   \multirow{2}{*}{\textbf{Model}} & \multirow{2}{*}{\textbf{\# Blocks}} & \multicolumn{3}{c}{\textbf{AUC}} \\
    & & \textbf{ML-1M} & \textbf{ML-20M} & \textbf{AMZ-Books} \\
    \midrule
    \multirow{6}{*}{HSTU} 
    & 2  & 0.7759 ($\uparrow$) & 0.7326 ($\downarrow$) & 0.7363 ($\uparrow$) \\
    & 4  & 0.7866 ($\uparrow$) & 0.7623 ($\downarrow$) & 0.7495 ($\uparrow$) \\
    & 8  & 0.7772 ($\uparrow$) & 0.7714 ($\downarrow$) & 0.7511 ($\uparrow$) \\
    & 16 & 0.7910 ($\downarrow$) & 0.7697 ($\downarrow$) & 0.7488 ($\uparrow$) \\
    & 24 & 0.7882 ($\uparrow$) & 0.7728 ($\downarrow$) & 0.7609 ($\uparrow$) \\
    & 32 & 0.7872 ($\downarrow$) & 0.7660 ($\downarrow$) & 0.7611 ($\uparrow$) \\
    \bottomrule
    \end{tabular}
    \caption{Performance of HSTU for the ranking task with a smaller item embedding size. The arrows indicate the direction of change in AUC compared to Table~\ref{tab:ranking}: ``$\uparrow$'' indicates an increase and ``$\downarrow$'' indicates a decrease.}
    \vspace{-7mm}
    \label{tab:small_dims}
\end{table}

\subsubsection{The Impact of Scoring Network Architecture}

\begin{figure}[t]
    \centering
    \includegraphics[width=0.95\columnwidth]{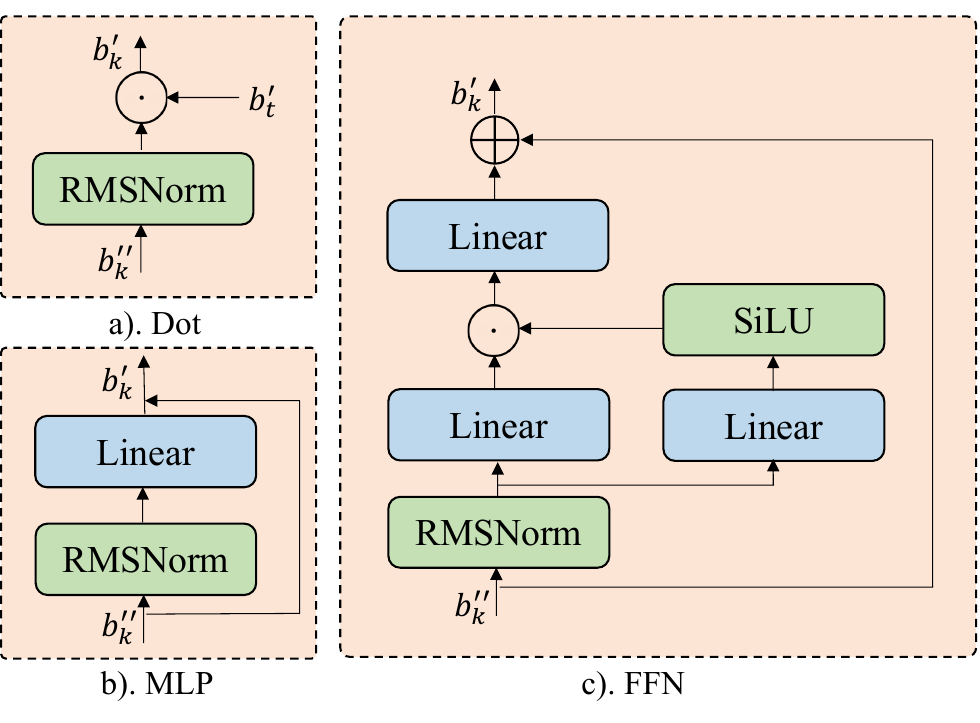}
    \vspace{-3mm}
    \caption{Illustration of scoring network architectures.}
    \label{fig:scoring_architecture}
\end{figure}

To further examine the impact of scoring network architecture, we employ a small neural network to generate scores for the ranking stage, utilizing the output from the models. To assess the impact of different neural network architectures, we implement three variations: (a) dot product (Dot), (b) multi-layer perceptron (MLP), and (c) Feed-forward networks (FFN). The detailed architectures of the neural networks are illustrated in Figure~\ref{fig:scoring_architecture}, where $b_t'$ denotes the target label. Among these, the dot product structure represents the simplest architecture, while the FFN is the most complex. 

The experimental results, presented in Table~\ref{tab:ranking_architecture}, reveal that for smaller datasets such as ML-1M and AMZ-Books, simpler scoring network architectures result in better model performance. Conversely, for larger datasets like ML-20M, more complex architectures yield superior performance. Larger datasets are generally better suited to more complex neural network architectures. These findings underscore the principle that stronger model capabilities do not universally translate to better performance; instead, optimal results are achieved when the architecture is appropriately tailored to the dataset size and complexity.

\subsubsection{Model Performance for Reduced Embedding Size}

As discussed in the previous section, achieving optimal recommendation performance requires aligning the model's expressive capability with the dataset size. For smaller datasets, increasing the model's expressive capability does not always result in better performance. To further investigate this finding, we conduct experiments using a very small embedding size of 4. This contrasts with the embedding sizes used in previous experiments, which are 50 for ML-1M, 256 for ML-20M, and 64 for AMZ-Books. The results of these experiments are presented in Table~\ref{tab:small_dims}.

Interestingly, when comparing the results with larger embedding sizes in Table~\ref{tab:ranking}, reducing the embedding size to a very small value improves performance for the smaller datasets, ML-1M and AMZ-Books. However, for the larger dataset, ML-20M, performance consistently declines. This observation indicates that for small datasets, using an excessively large embedding size may not be necessary and could even negatively impact model performance. 

At the same time, as shown in Table~\ref{tab:small_dims}, a smaller embedding size results in poorer scaling laws for ML-20M, despite the model's reduced parameter size compared to those in Table~\ref{tab:ranking}. This highlights an often-overlooked aspect: the scaling laws for generative recommendation models in ranking tasks are influenced not only by the vertical expansion (i.e., blocks of the transformer) of the model but also by its horizontal scale (i.e., embedding size). Previous studies on model parameter scaling have largely ignored the exploration of horizontal expansion.

Overall, Table~\ref{tab:small_dims} supports our conclusion that model size should match dataset size, with larger datasets requiring larger models. This finding leads to two key insights for future research directions:

\begin{itemize}[leftmargin=*,align=left]
\item In practical applications, the complexity and volume of user interaction data require more sophisticated models, highlighting the enhanced application potential and research value of generative large recommendation models that have scaling laws.

\item Parameter tuning for different datasets to achieve optimal performance can be resource-intensive. Future research should explore strategies for determining the optimal horizontal and vertical scaling of models (including embedding size and the number of layers) tailored to different datasets, to streamline the process and enhance recommendation performance.
\end{itemize}

\section{Future Directions}
\label{sec:future-directions}

In this paper, we have systematically reviewed the recent advancements in large recommendation models.
Our experiments involve both preliminary explorations and extensive evaluations using a combination of public datasets. These efforts highlight the significant potential of large recommendation models for future research and practical applications.
Despite these promising findings, it is important to acknowledge that the field of large recommendation models is still in its nascent stages. Our paper represents only an initial foray into this area, and further investigation is necessary to deepen understanding and enhance these models.

To aid future research endeavors, we have also identified and summarized several key perspectives from a substantial body of prior research that were not addressed in this paper. These insightful perspectives can help researchers pinpoint critical issues and pain points, thus offering promising opportunities for future exploration. Specifically, these perspectives primarily include unresolved challenges in data engineering, the application of tokenizers, and training/inference efficiency.
 
\subsection{Data Engineering}
The advent of deep learning has spurred the development of novel model-based recommendation methods aimed at enhancing performance. Despite these advancements, such methods often neglect a critical aspect: the analysis and quality improvement of the underlying recommendation data. A previous study~\cite{chin2022datasets} has demonstrated that the intrinsic characteristics of different recommendation datasets significantly influence the outcomes and comparative analyses of recommendation systems.

Recognizing this challenge, researchers are increasingly aware of the pivotal role of data quality in recommendation systems, which has led to the rise of data-centric approaches. 
The core principle of these approaches is that the quality and characteristics of the dataset fundamentally constrain model performance~\cite{lai2024survey,yin2024entropy}. 
To address data-related limitations, solutions such as data synthesis~\cite{wu2020joint, wang2019enhancing, huang2021mixgcf, yin2024dataset} and data denoising~\cite{quan2023robust, zhang2022hierarchical, lin2023autodenoise, wang2021denoising} have been proposed.

In parallel, the rapid development of large language models (LLMs) has showcased their exceptional capabilities across various natural language tasks. Researches~\cite{kaplan2020scaling, hoffmann2022empirical} highlight that the performance of LLMs consistently improves with increases in both the scale of training data and model size. This has shifted attention towards data engineering within LLMs, focusing on aspects such as data composition, quality control, and quantity~\cite{wang2024datamanagementtraininglarge}.

Given the critical role of data in both traditional recommendation models and LLMs, we posit that data engineering represents a vital research avenue in the realm of large recommendation models. Specifically, two key research directions emerge: quantifying the impact of data scale on model performance and defining as well as enhancing the quality of recommendation datasets. These areas hold significant potential for advancing the field.

\subsection{Tokenizer Application}

Tokenizers~\cite{li2023text, radford2021learning, ramesh2021zero, jia2021scaling, van2017neural, lee2022autoregressive} are fundamental components in deep learning, serving as a bridge between raw data and models. They play a crucial role in both recommendation systems and natural language processing (NLP) by enhancing the efficiency and scalability of deep learning solutions.

In recommendation systems, tokenizers convert user and item IDs into meaningful embedded representations, effectively capturing user preferences and item features. This transformation has spurred the development of various encoding techniques, including ID-based~\cite{koren2008factorization, he2017ncf, li2021sinkhorn}, text-based~\cite{zheng2017joint, kim2016convolutional, chen2018neural, catherine2017transnets, chin2018anr}, graph-based~\cite{grover2016node2vec, kipf2016variational, he2020lightgcn}, and compression encoding techniques~\cite{weinberger2009feature, jegou2010product}. These methods provide practical advantages in processing text data efficiently and enriching semantic information.

In the development of large language models (LLMs), tokenization technology has progressed from simple rule-based methods~\cite{rajaraman2024toward, brown1992class, koushik2019automated, yamout2018improved, yuan2018sentiment, mikolov2013efficient, pennington2014glove} to sophisticated context-aware models~\cite{sennrich2015neural, schuster2012japanese, kudo2018subword}. These advancements enhance the accuracy of language understanding and generation, while also broadening the potential for integrating diverse data modalities.

Given the critical role of tokenizers in recommendation systems and large models, they are poised to become even more crucial in large recommendation systems. This is particularly relevant when expanding recommendation datasets, as data expansion leads to a rapid increase in vocabulary size. The unique, dynamic, and extensive vocabulary in recommendation scenarios underscores the importance of tokenizers. We anticipate significant research opportunities in this area, focusing on the development of efficient and lossless tokenizers. Such research aims to minimize context information loss, improve processing speed and efficiency, and design tokenization strategies tailored to different modality features, thereby reducing text dependency. These studies hold promise for enhancing the capability and efficiency of tokenizers.

\subsection{Training/Inference Efficiency}

As data and parameters continue to expand, the computational and storage demands increase, creating a significant bottleneck in scaling models due to training and inference inefficiencies.

To address these challenges, researchers have explored various methods to scale LLMs. Techniques such as data parallelism~\cite{sergeev2018horovod, rasley2020deepspeed} and model parallelism~\cite{li2014scaling, wang2022merlin, guo2021scalefreectr, ivchenko2022torchrec} are used to efficiently handle large data volumes and model requests efficiently. Model structure compression techniques~\cite{zhao2024atom, hooper2024kvquant,liang2023less,huang2022context,sun2023simple, ma2023llm} simplify models at the algorithmic level, computational graph reconstruction methods~\cite{chen2018tvm, dao2023flashattention, zhai2023bytetransformer, rasley2020deepspeed} enhance efficiency during compilation, and system optimizations~\cite{kwon2023efficient, githubGitHubModelTClightllm, miao2023specinfer, jin2023s, kwon2023efficient, agrawal2024taming} improve overall throughput and reduce latency.

While there has been progress in optimizing the efficiency of general LLMs, the recommendation domain has seen limited exploration~\cite{2021Binary,wu2021linear,tian2023directed,wang2022compressing,yang2022adasparse, yu2022orca}, particularly concerning emerging large recommendation models. Recommendation systems face unique challenges, such as feature sparsity and the massive scale of users and items, resulting in a daily token processing volume that can be several orders of magnitude larger than what general LLMs handle over months. This imposes a substantial training burden and necessitates stringent real-time inference requirements.

Improving the training and inference efficiency and throughput of recommendation systems is thus a crucial research direction. Specific focus areas include achieving targeted software and hardware co-optimization based on the unique characteristics of recommendation systems, efficiently managing streaming data for parameter updates, and deploying systems with low resource requirements while ensuring efficient and accurate inference.
\section{Conclusion}
\label{sec:conclusion}

The emergence of large language models heightened interest in the scalability of models within the research community. The discovery of scaling laws had significant implications for recommendation systems as well. While Meta introduced HSTU and observed scaling law, several questions about scaling laws in large recommendation models remained unresolved. In this paper, we investigated the potential of large recommendation models, especially HSTU, across various recommendation tasks and the scaling laws they exhibited.
Firstly, we implemented a range of transformer-based architectures for large recommendation models and assessed their performance as model parameters were scaled up. We then conducted ablation studies to identify key modules that influenced scaling law. Furthermore, we examined the application potential of HSTU across different recommendation tasks. Additionally, we explored, for the first time, the performance and scaling laws of HSTU in ranking tasks. We believe this paper will shed light on future research regarding large recommendation models.
\balance

\bibliographystyle{ACM-Reference-Format}
\bibliography{sample-bibliography}
\end{document}